\documentclass[preprint,12pt]{article}

\usepackage{amsmath,
amssymb,
array,
calc,
rotating,
epsfig,
psfrag, 
amscd, 
datetime, 
comment, 
color, 
mathrsfs, 
ifthen,
caption,
subcaption}

\usepackage{marvosym}
\usepackage{tikzsymbols}
\usepackage[left=2cm,top=2cm,right=2cm]{geometry}
\usepackage{cite}

\usepackage{simpler-wick}
\usepackage{graphicx}

\numberwithin{equation}{section}

\newcommand{\nc}{\newcommand}

\definecolor{cardinal}{rgb}{0.6,0,0}
\definecolor{darkgreen}{rgb}{0,0.5,0}
\definecolor{golden}{rgb}{0.92, 0.7, 0}
\definecolor{midnight}{rgb}{0, 0, 0.5}
\definecolor{darkblue}{rgb}{0.2, 0, 0.8}

\usepackage{hyperref}

\nc{\ra}{\rightarrow} 
\nc{\lra}{\leftrightarrow} 
\nc{\Ra}{\Rightarrow} 
\nc{\LRa}{\Leftightarrow} 
\nc{\blp}{{\big (}}
\nc{\brp}{{\big )}}
\nc{\Blp}{{\Big (}}
\nc{\Brp}{{\Big )}}
\nc{\bglp}{{\bigg (}}
\nc{\bgrp}{{\bigg )}}
\nc{\Bglp}{{\Bigg (}}
\nc{\Bgrp}{{\Bigg )}}
\nc{\slb}{{\rm [}}
\nc{\srb}{{\rm ]}}
\nc{\bslb}{{\rm \big [}}
\nc{\bsrb}{{\rm \big ]}}
\nc{\Bslb}{{\rm \Big [}}
\nc{\Bsrb}{{\rm \Big ]}}

\def\al{\alpha}

\def\eps{\epsilon}
\nc{\veps}{\varepsilon}
\def\gam{\gamma}

\def\lam{\lambda}
\def\om{\omega}

\nc{\vphi}{\varphi}
\def\tha{\theta}

\def\sig{\sigma}

\def\Gam{\Gamma}

\def\Lam{\Lambda}
\def\Om{\Omega}
\def\Sig{\Sigma}

\def\coeff#1#2{\relax{\textstyle {#1 \over #2}}\displaystyle}

\nc{\myvspace}{\rule[-1em]{0pt}{2.5em}}
\nc{\bea}{\begin{eqnarray}}
\nc{\eea}{\end{eqnarray}}
\nc{\be}{\begin{equation}}
\nc{\ee}{\end{equation}}
\nc{\barr}{\begin{array}}
\nc{\earr}{\end{array}}

\nc{\co}{{\cal o}}

\nc{\cA}{{\cal A}}
\nc{\cB}{{ \cal B}}

\def\cD{{\cal D}}

\nc{\cF}{{\cal F}}
\nc{\cG}{{\cal G}}

\def\cI{{\cal I}}

\def\cK{{\cal K}}
\nc{\cL}{{\cal L}}
\nc{\cM}{{\cal M}}

\def\cN{{\cal N}}

\def\cO{{\cal O}}

\nc{\cQ}{{\cal Q}}
\nc{\cR}{{\cal R}}
\def\cS{{\cal S}}

\def\cV{{\cal V}}
\def\cV{{\cal V}}
\def\cW{{\cal W}}

\def\cZ{{\cal Z}}
\nc{\cQd}{\cQ^{\dagger}}
\nc{\cRd}{\cR^{\dagger}}
\nc{\BB}{{\mathbb B}}
\nc{\CC}{{\mathbb C}}
\nc{\DD}{{\mathbb D}}
\nc{\EE}{{\mathbb E}}
\nc{\FF}{{\mathbb F}}
\nc{\GG}{{\mathbb G}}
\nc{\HH}{{\mathbb H}}
\nc{\JJ}{{\mathbb J}}
\nc{\MM}{{\mathbb M}}
\nc{\RR}{{\mathbb R}}
\nc{\PP}{{\mathbb P}}
\nc{\QQ}{{\mathbb Q}}
\nc{\UU}{{\mathbb U}}
\nc{\ZZ}{{\mathbb Z}}
\nc{\calone}{{\mathbb 1}}

\nc{\half}{\coeff{1}{2}}
\nc{\quarter}{\coeff{1}{4}}
\nc{\del}{\partial}

\nc{\delbar}{\bar\partial}
\nc{\thalf}{\frac{t}{2}}
\nc{\Spin}{\operatorname{Spin}}
\nc{\SO}{\operatorname{SO}}

\nc{\Sp}{{\rm Sp}}
\nc{\com}[2]{{ \left[ #1, #2 \right] }}
\nc{\acom}[2]{{ \left\{ #1, #2 \right\} }}
\nc{\rr}{\rightarrow}
\nc{\p}{\partial}
\nc{\LT}{{\LL_\T}}
\nc{\Tr}{{\rm Tr}}
\nc{\tr}{{\rm tr}}
\nc{\Adag}{A^{\dagger}}
\nc{\AdagI}{A^{\dagger I}}
\nc{\AdagJ}{A^{\dagger J}}
\nc{\AdagK}{A^{\dagger K}}
\nc{\AdagL}{A^{\dagger L}}
\nc{\AdagM}{A^{\dagger M}}
\nc{\Bdag}{B^{\dagger}}
\nc{\BdagI}{B^{\dagger}_I}
\nc{\BdagJ}{B^{\dagger}_J}
\nc{\BdagK}{B^{\dagger}_K}
\nc{\BdagL}{B^{\dagger}_L}
\nc{\BdagM}{B^{\dagger}_M}
\nc{\Cdag}{C^{\dagger}}
\nc{\CdagI}{C^{\dagger I}}
\nc{\CdagJ}{C^{\dagger J}}
\nc{\CdagK}{C^{\dagger K}}
\nc{\Ddag}{D^{\dagger}}
\nc{\DdagI}{D^{\dagger I}}
\nc{\DdagJ}{D^{\dagger J}}
\nc{\DdagK}{D^{\dagger K}}
\nc{\bva}{\breve{a}}
\nc{\bvb}{\breve{b}}
\nc{\bvc}{\breve{c}}
\nc{\bvd}{\breve{d}}
\nc{\bve}{\breve{e}}
\nc{\bvf}{\breve{f}}
\nc{\bvg}{\breve{g}}
\nc{\bvh}{\breve{h}}
\nc{\bvi}{\breve{i}}
\nc{\bvj}{\breve{j}}
\nc{\bvk}{\breve{k}}
\nc{\bvl}{\breve{l}}
\nc{\bvm}{\breve{m}}
\nc{\bvn}{\breve{n}}
\nc{\bvo}{\breve{o}}
\nc{\bvp}{\breve{p}}
\nc{\brvq}{\breve{q}}
\nc{\bvr}{\breve{r}}
\nc{\bvs}{\breve{s}}
\nc{\bvt}{\breve{t}}
\nc{\bvu}{\breve{u}}
\nc{\bvv}{\breve{v}}
\nc{\bvw}{\breve{w}}
\nc{\bvx}{\breve{x}}
\nc{\bvy}{\breve{y}}
\nc{\bvz}{\breve{z}}

\nc{\bvA}{\breve{A}}
\nc{\bvB}{\breve{B}}
\nc{\bvC}{\breve{C}}
\nc{\bvD}{\breve{D}}
\nc{\bvE}{\breve{E}}
\nc{\bvF}{\breve{F}}
\nc{\bvG}{\breve{G}}
\nc{\bvH}{\breve{H}}
\nc{\bvI}{\breve{I}}
\nc{\bvJ}{\breve{J}}
\nc{\bvK}{\breve{K}}
\nc{\bvL}{\breve{L}}
\nc{\bvM}{\breve{M}}
\nc{\bvN}{\breve{N}}
\nc{\bvO}{\breve{O}}
\nc{\bvP}{\breve{P}}
\nc{\bvQ}{\breve{Q}}
\nc{\bvR}{\breve{R}}
\nc{\bvS}{\breve{S}}
\nc{\bvT}{\breve{T}}
\nc{\bvU}{\breve{U}}
\nc{\bvV}{\breve{V}}
\nc{\bvcV}{\breve{\cV}}
\nc{\bvW}{\breve{W}}
\nc{\bvX}{\breve{X}}
\nc{\bvY}{\breve{Y}}
\nc{\bvZ}{\breve{Z}}

\nc{\ul}[1]{{\underline{#1}}}

\nc{\tal}{\widetilde{\alpha}}
\nc{\tbeta}{\widetilde{\beta}}
\nc{\ttha}{\tilde{\theta}}
\nc{\ttau}{\tilde{\tau}}
\nc{\tTha}{\tilde{\Theta}}
\nc{\tphi}{\tilde{\phi}}
\nc{\tsig}{\tilde{\sig}}
\nc{\tom}{\widetilde{\om}}
\nc{\tOm}{\widetilde{\Om}}
\nc{\tlam}{\widetilde{\lam}}
\nc{\tLam}{\tilde{\Lam}}
\nc{\tSig}{\widetilde{\Sig}}
\nc{\tPhi}{\tilde{\Phi}}
\nc{\tPhibar}{\ol{\tPhi}}
\nc{\tPi}{\widetilde{\Pi}}
\nc{\tpsi}{\widetilde{\psi}}
\nc{\tPsi}{\tilde{\Psi}}
\nc{\tgam}{\widetilde{\gam}}
\nc{\tGam}{\widetilde{\Gam}}
\nc{\tzeta}{\tilde{\zeta}}
\nc{\tZeta}{\tilde{\Zeta}}
\nc{\teta}{\widetilde{\eta}}
\nc{\teps}{\tilde{\eps}}
\nc{\tveps}{\tilde{\veps}}
\nc{\tEta}{\tilde{\Eta}}
\nc{\tchi}{\tilde{\chi}}
\nc{\tChi}{\tilde{\Chi}}
\nc{\txi}{\tilde{\xi}}
\nc{\tXi}{\widetilde{\Xi}}
\nc{\tnu}{\tilde{\nu}}
\nc{\tmu}{\tilde{\mu}}

\nc{\ta}{\tilde a}
\nc{\tb}{\tilde b}
\nc{\tc}{\tilde c}
\nc{\te}{\tilde e}
\nc{\tf}{\widetilde f}
\nc{\tg}{\widetilde g}
\nc{\ti}{\tilde i}
\nc{\tj}{\tilde j}
\nc{\tk}{\widetilde k}
\nc{\tl}{\tilde l}
\nc{\tm}{\widetilde m}
\nc{\tn}{\tilde n}
\nc{\tp}{\tilde{p}}
\nc{\tq}{\widetilde{q}}
\nc{\trr}{{\tilde r}}
\nc{\ts}{{\tilde s}}
\nc{\tu}{{\tilde u}}
\nc{\tv}{{\tilde v}}
\nc{\tw}{{\tilde w}}
\nc{\tx}{{\tilde x}}
\nc{\ty}{{\tilde y}}
\nc{\tz}{\tilde z}
\nc{\tA}{{\widetilde A}}
\nc{\tAbar}{{\ol \tA}}
\nc{\tB}{{\widetilde B}}
\nc{\tC}{{\widetilde C}}
\nc{\tD}{{\widetilde D}}
\nc{\tE}{{\widetilde E}}
\nc{\tF}{{\widetilde F}}
\nc{\tG}{{\widetilde G}}
\nc{\tcG}{{\widetilde \cG}}
\nc{\tH}{{\widetilde H}}
\nc{\tI}{{\widetilde I}}
\nc{\tcI}{{\widetilde \cI}}
\nc{\tJ}{{\widetilde J}}
\nc{\tJbar}{{\ol {\tilde J}}}
\nc{\tK}{{\widetilde K}}
\nc{\tL}{{\widetilde L}}
\nc{\tcL}{{\widetilde \cL}}
\nc{\tcLbar}{{\ol \tcL}}
\nc{\tM}{{\widetilde M}}
\nc{\tN}{{\widetilde N}}
\nc{\tcN}{{\widetilde \cN}}
\nc{\tcO}{{\widetilde \cO}}
\nc{\tO}{{\widetilde O}}
\nc{\tP}{{\widetilde P}}
\nc{\tQ}{{\widetilde Q}}
\nc{\tR}{{\widetilde R}}
\nc{\tS}{\widetilde{S}}
\nc{\tT}{\widetilde{T}}
\nc{\tU}{\widetilde{U}}
\nc{\tUU}{\widetilde{\UU}}
\nc{\tV}{\widetilde{V}}
\nc{\tcV}{\widetilde{\cV}}
\nc{\tcVbar}{\ol{\widetilde{\cV}}}
\nc{\tW}{\widetilde{W}}
\nc{\tcF}{\widetilde{{\cal F}}}
\nc{\tX}{\widetilde{X}}
\nc{\tY}{\widetilde{Y}}
\nc{\tcZ}{\tilde{\cZ}}
\nc{\tcZbar}{\ol{\tcZ}}

\nc{\ha}{\hat a}
\nc{\hb}{\hat b}
\nc{\hc}{\widehat c}
\nc{\hd}{\widehat d}
\nc{\he}{\widehat e}
\nc{\hf}{\widehat f}
\nc{\hg}{\widehat g}
\nc{\hh}{\widehat h}
\nc{\hm}{\widehat m}
\nc{\hn}{\widehat n}
\nc{\hp}{\widehat p}
\nc{\hq}{\widehat q}
\nc{\hr}{\widehat r}
\nc{\hs}{\widehat s}
\nc{\hv}{\widehat v}
\nc{\hw}{\widehat w}
\nc{\hx}{\widehat x}
\nc{\hy}{\widehat y}
\nc{\hz}{\widehat z}
\nc{\zhat}{\hat z}
\nc{\hA}{\widehat{A}}
\nc{\hB}{\widehat{B}}
\nc{\hC}{\widehat{C}}
\nc{\hD}{\widehat{D}}
\nc{\hE}{\widehat{E}}
\nc{\hF}{\widehat{F}}
\nc{\hcF}{\widehat{\cF}}
\nc{\hG}{\widehat{G}}
\nc{\hcG}{\widehat{\cG}}
\nc{\hH}{\widehat{H}}
\nc{\hI}{\widehat{I}}
\nc{\hcI}{\widehat{\cI}}
\nc{\hJ}{\widehat{J}}
\nc{\hK}{\widehat{K}}
\nc{\hL}{\widehat{L}}
\nc{\hcL}{\widehat{\cL}}
\nc{\hM}{\widehat M}
\nc{\hcM}{\widehat{\cM}}
\nc{\hN}{\widehat{N}}
\nc{\hO}{\widehat{O}}
\nc{\hcO}{\widehat{\cO}}
\nc{\hP}{\widehat{P}}
\nc{\hQ}{\widehat{Q}}
\nc{\hcQ}{\widehat{\cQ}}
\nc{\hcR}{\widehat{\cR}}
\nc{\hR}{\widehat{R}}
\nc{\hS}{\widehat{S}}
\nc{\hcS}{\widehat{\cS}}
\nc{\hT}{\widehat{T}}
\nc{\hU}{\widehat{U}}
\nc{\hV}{\widehat V}
\nc{\hcV}{\widehat \cV}
\nc{\hX}{\widehat X}
\nc{\hcZ}{\widehat \cZ}
\nc{\hcZbar}{\ol{\widehat \cZ}}

\nc{\heta}{\widehat{\eta}}
\nc{\hal}{\widehat \alpha}
\nc{\hbeta}{\widehat \beta}
\nc{\hphi}{\widehat{\phi}}
\nc{\hkap}{\hat{\kappa}}
\nc{\hchi}{\widehat{\chi}}
\nc{\hpsi}{\widehat{\psi}}
\nc{\hgam}{\widehat{\gam}}
\nc{\hPhi}{\hat{\Phi}}
\nc{\hPsi}{\hat{\Psi}}
\nc{\hGam}{\hat{\Gam}}
\nc{\omhat}{\widehat{\om}}
\nc{\htha}{\hat{\tha}}
\nc{\hrho}{\widehat{\rho}}
\nc{\hdel}{\widehat{\del}}
\nc{\hnabla}{\widehat{\nabla}}

\nc{\w}{\wedge}

\nc{\vb}{\vec b}
\nc{\vc}{\vec c}
\nc{\vd}{\vec d}
\nc{\ve}{\vec e}
\nc{\vf}{\vec f}
\nc{\vg}{\vec g}
\nc{\vh}{\vec h}
\nc{\vp}{\vec p}
\nc{\vq}{\vec q}
\nc{\vr}{\vec r}
\nc{\vs}{\vec s}
\nc{\vv}{\vec v}
\nc{\vw}{\vec w}
\nc{\vx}{\vec x}
\nc{\vy}{\vec y}
\nc{\vz}{\vec z}

\nc{\vB}{\vec B}
\nc{\vC}{\vec C}
\nc{\vD}{\vec D}
\nc{\vE}{\vec E}
\nc{\vF}{\vec F}
\nc{\vG}{\vec G}
\nc{\vH}{\vec H}
\nc{\vP}{\vec P}
\nc{\vQ}{\vec Q}
\nc{\vR}{\vec R}
\nc{\vS}{\vec S}
\nc{\vV}{\vec V}
\nc{\vW}{\vec W}
\nc{\vX}{\vec X}
\nc{\vY}{\vec Y}
\nc{\vZ}{\vec Z}

\nc{\ol}{\overline}
\nc{\abar}{\ol{a}}
\nc{\bbar}{\ol{b}}
\nc{\cbar}{\ol{c}}
\nc{\dbar}{\ol{d}}
\nc{\ebar}{\ol{e}}
\nc{\fbar}{\ol{f}}
\nc{\gbar}{\ol{g}}
\nc{\ibar}{\ol{\imath}}
\nc{\jbar}{\ol{\jmath}}
\nc{\kbar}{\ol{k}}
\nc{\lbar}{\ol{l}}
\nc{\mbar}{\ol{m}}
\nc{\nbar}{\ol{n}}
\nc{\pbar}{\ol{p}}
\nc{\qbar}{\ol{q}}
\nc{\rbar}{\ol{r}}
\nc{\sbar}{\ol{s}}
\nc{\ubar}{\ol{u}}
\nc{\vbar}{\ol{v}}
\nc{\wbar}{\ol{w}}
\nc{\xbar}{\ol{x}}
\nc{\ybar}{\ol{y}}
\nc{\zbar}{\ol{z}}

\nc{\Abar}{\ol{A}}
\nc{\Bbar}{\ol{B}}
\nc{\cBbar}{\ol{\cB}}
\nc{\Cbar}{\ol{C}}
\nc{\Dbar}{\ol{D}}
\nc{\Ebar}{\ol{E}}
\nc{\Fbar}{\ol{F}}
\nc{\Gbar}{\ol{G}}
\nc{\Jbar}{\ol{J}}
\nc{\Kbar}{\ol{K}}
\nc{\cKbar}{\ol{\cK}}
\nc{\Lbar}{\ol{L}}
\nc{\cLbar}{\ol{\cL}}
\nc{\Mbar}{\ol{M}}
\nc{\Nbar}{\ol{N}}
\nc{\Pbar}{\ol{P}}
\nc{\Qbar}{\ol{Q}}
\nc{\Rbar}{\ol{R}}
\nc{\Sbar}{\ol{S}}
\nc{\Tbar}{\ol{T}}
\nc{\Ubar}{\ol{U}}
\nc{\Vbar}{\ol{V}}
\nc{\cVbar}{\ol{\cV}}
\nc{\Wbar}{\ol{W}}
\nc{\cWbar}{\ol{\cW}}
\nc{\Xbar}{{\overline X}}
\nc{\Ybar}{{\overline Y}}
\nc{\Zbar}{{\overline Z}}
\nc{\cZbar}{{\overline \cZ}}

\nc{\epsbar}{\ol{\epsilon}}
\nc{\albar}{\ol{\al}}
\nc{\Albar}{\ol{\Al}}
\nc{\betabar}{\ol{\beta}}
\nc{\Betabar}{\ol{\Beta}}
\nc{\lambar}{\ol{\lambda}}
\nc{\kapbar}{\ol{\kappa}}
\nc{\zetabar}{\ol{\zeta}}
\nc{\Zetabar}{\ol{\Zeta}}
\nc{\taubar}{\ol{\tau}}
\nc{\Taubar}{\ol{\Tau}}
\nc{\psibar}{\ol{\psi}}
\nc{\Psibar}{\ol{\Psi}}
\nc{\tpsibar}{\ol{\tpsi}}
\nc{\tPsibar}{\ol{\tPsi}}
\nc{\phibar}{\ol{\phi}}
\nc{\Phibar}{\ol{\Phi}}
\nc{\chibar}{\ol{\chi}}
\nc{\mubar}{\ol{\mu}}
\nc{\nubar}{\ol{\nu}}
\nc{\rhobar}{\ol{\rho}}
\nc{\ombar}{\ol{\om}}
\nc{\Ombar}{\ol{\Om}}
\nc{\Deltabar}{\ol{\Delta}}
\nc{\Thetabar}{\ol{\Theta}}
\nc{\xibar}{\ol{\xi}}
\nc{\Xibar}{\ol{\Xi}}

\nc{\Dthbar}{\ol{\rm D3}}

\nc{\fdot}{\dot{f}}
\nc{\gdot}{\dot{g}}
\nc{\pdot}{\dot{p}}
\nc{\qdot}{\dot{q}}
\nc{\rdot}{\dot{r}}
\nc{\sdot}{\dot{s}}
\nc{\tdot}{\dot{t}}
\nc{\udot}{\dot{u}}
\nc{\vdot}{\dot{v}}
\nc{\wdot}{\dot{w}}
\nc{\xdot}{\dot{x}}
\nc{\xddot}{\ddot{x}}
\nc{\ydot}{\dot{y}}
\nc{\zdot}{\dot{z}}
\nc{\yddot}{\ddot{y}}

\nc{\Adot}{\dot{A}}
\nc{\Bdot}{\dot{B}}
\nc{\Cdot}{\dot{C}}
\nc{\Udot}{\dot{U}}
\nc{\Vdot}{\dot{V}}
\nc{\Wdot}{\dot{W}}

\nc{\taudot}{\dot{\tau}}
\nc{\phidot}{\dot{\phi}}
\nc{\psidot}{\dot{\psi}}
\nc{\chidot}{\dot{\chi}}
\nc{\sinp}{s_{\phi}}
\nc{\cosp}{c_{\phi}}
\nc{\tanp}{t_{\phi}}
\nc{\spone}{s_{\phi_1}}
\nc{\cpone}{c_{\phi_1}}
\nc{\tpone}{t_{\phi_1}}
\nc{\sptwo}{s_{\phi_2}}
\nc{\cptwo}{c_{\phi_2}}
\nc{\tptwo}{t_{\phi_2}}
\nc{\spth}{s_{\phi_3}}
\nc{\cpth}{c_{\phi_3}}
\nc{\tpth}{t_{\phi_3}}
\nc{\calp}{c_{\al}}
\nc{\salp}{s_{\al}}

\nc{\csch}{{\rm csch}}
\nc{\sech}{{\rm sech}}

\nc{\cothzlami}{\coth(z-\lam_i)}
\nc{\coshzlami}{\cosh(z-\lam_i)}
\nc{\sinhzlami}{\sinh(z-\lam_i)}

\nc{\cothzlamj}{\coth(z-\lam_j)}
\nc{\coshzlamj}{\cosh(z-\lam_j)}
\nc{\sinhzlamj}{\sinh(z-\lam_j)}

\nc{\cothlamij}{\coth(\lam_i-\lam_j)}
\nc{\coshlamij}{\cosh(\lam_i-\lam_j)}
\nc{\sinhlamij}{\sinh(\lam_i-\lam_j)}

\nc{\bah}{{\mathbf {\hat{A}}}}
\nc{\bX}{{\mathbf X}}
\nc{\ba}{{\bf a}}
\nc{\bb}{{\bf b}}
\nc{\bc}{{\bf c}}
\nc{\bd}{{\bf d}}
\nc{\bg}{{\bf g}}
\nc{\bk}{{\bf k}}
\nc{\bl}{{\bf l}}
\nc{\bm}{{\bf m}}
\nc{\bn}{{\bf n}}
\nc{\bo}{{\bf o}}
\nc{\bp}{{\bf p}}
\nc{\bq}{{\bf q}}
\nc{\br}{{\bf r}}
\nc{\bs}{{\bf s}}
\nc{\bt}{{\bf t}}
\nc{\bu}{{\bf u}}
\nc{\bv}{{\bf v}}
\nc{\bw}{{\bf w}}
\nc{\bx}{{\bf x}}
\nc{\by}{{\bf y}}
\nc{\bz}{{\bf z}}

\nc{\bP}{{\bf P}}
\nc{\bQ}{{\bf Q}}

\nc{\bom}{{\bf \om}}
\nc{\bombar}{{\mathbf \ombar}}
\nc{\bPhi}{{\bf \Phi}}

\nc{\rma}{{\rm a}}
\nc{\rmb}{{\rm b}}
\nc{\rmc}{{\rm c}}
\nc{\rmd}{{\rm d}}
\nc{\rmg}{{\rm g}}
\nc{\rk}{{\rm k}}
\nc{\rml}{{\rm l}}
\nc{\rmm}{{\rm m}}
\nc{\rmn}{{\rm n}}
\nc{\rmo}{{\rm o}}
\nc{\rmp}{{\rm p}}
\nc{\rmq}{{\rm q}}
\nc{\rmr}{{\rm r}}
\nc{\rms}{{\rm s}}
\nc{\rmt}{{\rm t}}
\nc{\rmu}{{\rm u}}
\nc{\rmv}{{\rm v}}
\nc{\rmw}{{\rm w}}
\nc{\rmx}{{\rm x}}
\nc{\rmy}{{\rm y}}
\nc{\rmz}{{\rm z}}

\nc{\dal}{\dot{\al}}
\nc{\thadot}{\dot{\tha}}
\nc{\thab}{\bar{\theta}}
\nc{\thal}{\theta^{\al}}
\nc{\thdal}{\bar{\theta}^{\dal}}

\nc{\thsigthm}{\tha \sigma^m \thab}
\nc{\thsigthn}{\tha \sigma^n \thab}

\nc{\Dal}{D_{\al}}
\nc{\Ddal}{\bar{D}_{\dal}}
\nc{\CDal}{{\cal D}_{\al}}
\nc{\CDdal}{\bar{\cal D}_{\dal}}

\nc{\eq}[1]{{(\ref{#1})}}
\nc{\eqtwo}[2]{{(\ref{#1},\ref{#2})}}
\nc{\eqthree}[3]{(\ref{#1},\ref{#2},\ref{#3})}
\nc{\eqfour}[4]{(\ref{#1},\ref{#2},\ref{#3},\ref{#4})}
\nc{\eqfive}[5]{(\ref{#1},\ref{#2},\ref{#3},\ref{#4,\ref{#5}})}
\nc{\non}{\nonumber}
\nc{\Fzero}{F_{(0)}}
\nc{\Ftwo}{F_{(2)}}
\nc{\Ffour}{F_{(4)}}
\nc{\Fone}{F_{(1)}}
\nc{\Fthree}{F_{(3)}}
\nc{\Ffive}{F_{(5)}}
\nc{\Fn}{F_{(n)}}
\nc{\Fp}{F_{(p)}}

\nc{\tFzero}{\tF_{(0)}}
\nc{\tFtwo}{\tF_{(2)}}
\nc{\tFfour}{\tF_{(4)}}
\nc{\tFone}{\tF_{(1)}}
\nc{\tFthree}{\tF_{(3)}}
\nc{\tFfive}{\tF_{(5)}}
\nc{\tFn}{\tF_{(n)}}
\nc{\tFp}{\tF_{(p)}}

\nc{\Czero}{C_{(0)}}
\nc{\Ctwo}{C_{(2)}}
\nc{\Cfour}{C_{(4)}}
\nc{\Cone}{C_{(1)}}
\nc{\Cthree}{C_{(3)}}
\nc{\Cfive}{C_{(5)}}
\nc{\Cn}{C_{(n)}}

\nc{\IGGGG}{{I_4(\cG,\cG,\cG,\cG)}}
\nc{\IGGGQ}{{I_4(\cG,\cG,\cG,\cQ)}}
\nc{\IGGQQ}{{I_4(\cG,\cG,\cQ,\cQ)}}
\nc{\IGQQQ}{{I_4(\cG,\cQ,\cQ,\cQ)}}
\nc{\IQQQQ}{{I_4(\cQ,\cQ,\cQ,\cQ)}}

\nc{\IpGGG}{{I_4(\cG,\cG,\cG)}}
\nc{\IpGGQ}{{I_4(\cG,\cG,\cQ)}}
\nc{\IpGQQ}{{I_4(\cG,\cQ,\cQ)}}
\nc{\IpQQQ}{{I_4(\cQ,\cQ,\cQ)}}

\nc{\IGQ}{\langle \cG,\cQ \rangle}

\nc{\IGGGQQQ}{\langle \IpGGG,\IpQQQ \rangle}

\nc{\equ}{{\rm eq}}
\def\Im{{\rm Im \hspace{0.5mm} }}

\def\Re{{\rm Re \hspace{0.5mm}}}

\nc{\vol}{{\rm vol}}
\nc{\Ainf}{A_{\infty}}
\nc{\End}{{\rm End}}
\nc{\Ext}{{\rm Ext}}
\nc{\IIB}{{\rm IIB}}
\nc{\Ad}{{\rm Ad}}
\nc{\IIA}{{\rm IIA}}
\nc{\AdS}{{\rm AdS}}
\nc{\CFT}{{\rm CFT}}
\nc{\diag}{{\rm diag}}
\nc{\Log}{{\rm Log}}
\nc{\Dslash}{\ensuremath \raisebox{0.025cm}{\slash}\hspace{-0.32cm} D}
\nc{\cDslash}{\ensuremath \raisebox{0.025cm}{\slash}\hspace{-0.32cm} \cD}
\nc{\omslash}{\om\!\!\!/}
\nc{\no}{\!:\!\!}
\nc{\ointdz}{\oint\frac{dz}{2\pi i}}
\nc{\ointdzone}{\oint\frac{dz_1}{2\pi i}}
\nc{\ointdztwo}{\oint\frac{dz_2}{2\pi i}}
\nc{\ointdzb}{\oint\frac{d\zbar}{2\pi i}}
\nc{\ointdzbone}{\oint\frac{d\zbar_1}{2\pi i}}
\nc{\ointdzbtwo}{\oint\frac{d\zbar_2}{2\pi i}}
\nc{\dz}{\frac{dz}{2\pi i}}
\nc{\dzb}{\frac{d\zbar}{2\pi i}}
\nc{\bpm}{\begin{pmatrix}}
\nc{\epm}{\end{pmatrix}}
 \nc{\bitem}{\begin{itemize}}
 \nc{\eitem}{\end{itemize}}
 \nc{\exercise}{\vskip 2mm \noindent {\bf Exercise:}}
 \nc{\definition}{\vskip 2mm \noindent {\bf Definition:}}
 
\begin{document}

\vspace{0.5cm}
\begin{center}
\baselineskip=13pt {\LARGE \bf{Mixed Moments for the \\
Product of Ginibre Matrices}\\}
 \vskip1.5cm 
Nick Halmagyi$^{\BasicTree[1.5]{green!20!black}{green!50!black}{green!70!black}{leaf}}$, Shailesh Lal$^{\tikzsymbolsuse{drWalley}}$\ 

 \vskip0.5cm
${\BasicTree[1.5]{green!20!black}{green!50!black}{green!70!black}{leaf}}$
\textit{LPTHE, CNRS \& Sorbonne Universit\'e, \\
4 Place Jussieu, F-75252, Paris, France 
}\\
\vskip0.1cm
halmagyi@lpthe.jussieu.fr \\
\vskip0.8cm

${\tikzsymbolsuse{drWalley}}$
\textit{Faculdade de Ciencias, Universidade do Porto,\\
687 Rua do Campo Alegre, Porto 4169-007, Portugal.}\\
\vskip0.1cm
slal@fc.up.pt \\
\vskip3cm
\end{center}

\begin{abstract}
We study the ensemble of a product of $n$ complex Gaussian i.i.d. matrices. We find this ensemble is Gaussian with a variance matrix which is averaged over a multi-Wishart ensemble. We compute the mixed moments and find that at large $N$, they are given by an enumeration of non-crossing pairings weighted by Fuss-Catalan numbers.
\end{abstract}

\newpage
\section{Introduction}

The theory of random matrices (RMT) has a remarkably broad reach through physics, mathematics, biology, computer science and engineering. It is uncanny how fundamental contributions to RMT are regularly made by researchers in each of these domains. In the current work, with a view towards studying feedforward neural networks, we draw from results in these fields and study the mixed moments of products of complex i.i.d. Gaussian matrices\footnote{The archetypical result in this field is due to Ginibre \cite{Ginibre1965} who computed the eigenvalue distribution of these matrices and found it to be uniform on the disk, we will refer to complex i.i.d. Gaussian matrices as Ginibre matrices.}.

A defining feature of a Ginibre matrix, certainly in contrast to Hermitian matrices, is that they cannot be diagonalized; under the Schur decomposition of a complex matrix
\be
X = U (\Lam + T)U^{-1}\,,\qquad \Lam=\diag(\lam_1\,,\ldots \lam_N)\,,\quad  U\in U(N)  \label{SchurDecomp}
\ee
there is a non-trivial upper triangular component $T$ unless $X$ is Hermitian.
Recall that diagonalization is at the heart of many results in Hermitian RMT \cite{Brezin:1977sv}, it allows a problem which a priori has $O(N^2)$ degrees of freedom to be reduced to an $O(N)$ computation. Ginibre managed to circumvent this obstacle by exploiting the fact that in the computation of the eigenvalue spectrum, the components of $T$ are decoupled Gaussian degrees of freedom and can be integrated. However the presence of $O(N^2)$ coupled degrees of freedom is not necessarily something one should shy away from; while the Hermitian matrix models have a single operator for a fixed number of insertions, the number of operators in the Ginibre ensemble grows exponentially with the number of insertions. It is thus evident that the space of Ginibre matrices is far richer that the space of Hermitian matrices and one should expect crucial features to be unattainable through methods which reduce the problem to $O(N)$ degrees of freedom. 

There are numerous ways to generalize the Ginibre ensemble, two ideas which influenced the current work are:
\begin{enumerate}
\item Relax the i.i.d. condition while preserving the Gaussian structure of the distribution.
\item Consider products of Ginibre matrices.
\end{enumerate}
To some extent both these generalizations are designed to produce solvable random matrix ensembles while remaining within, or at least in the vicinity of, the safe harbours of Gaussian ensembles. However we will in fact find, that products of Ginibre matrices remain Gaussian but are no longer identically distributed. 

One way to relax explicitly the i.i.d. condition is to introduce a fixed variance matrix (or matrices), for example\footnote{
Another interesting way to relax the i.i.d condition intiated in \cite{PhysRevLett.60.1895} (see \cite{Khoruzhenko09115645} for a concise review), is to study Gaussian ensembles of the form
\be
P(X,X^\dagger) \sim e^{-\Tr XX^\dagger + (\tau \Tr XX+c.c)}
\ee
where $\tau$ is a complex scalar which one might call the {\it elliptical} parameter. 
}
\be
P_{\Sig}(X) \sim e^{-N \Tr X\Sig^{-1}X^\dagger}
\ee
and then when deploying Wick's theorem one uses the propogator
\be
\langle X_{ij}X_{kl}^{\dagger} \rangle = \frac{1}{N} \delta_{il} \Sig^{jk}\,.
\ee
This particular deformation of the Ginibre ensemble has the effect of decorating the graphs which contribute to the moments of $X,X^\dagger$ with a product of traces of powers of $\Sig$ and the precise enumeration is challenging. Nonetheless there are numerous results regarding computations of moments and limiting eigenvalue distributions reviewed in the nice books \cite{TulinoVerduBook, forrester2010log,  BaiSilverteinBook, Couillet:2011:RMM:2161679}, which we will use throughout this work.

The ensemble of products of random matrices has been an attractive research area for some time, much of the history is summarised nicely in the thesis \cite{Ipsen151006128}. In the last decade, there has however been a snowballing of results regarding this ensemble, yet strangely enough, this does not appear to be due to any particularly new technique which has come online\footnote{
An interesting more modern development in RMT is the development of the Schwinger-Dyson (loop) equations into topological recursion \cite{Eynard:2007kz} and there has been a recent work investigating such modern techniques to the product Ginibre ensemble \cite{dartois2019schwingerdyson}.
} but rather it has been motivated by various potential applications. In that vein, we were drawn to this topic through the study of simple models of feed forward neural networks \cite{Ganguli_exact} and while the techniques we use are not particularly new, we certainly draw upon many recent results. We will find that the probability distribution of the product of Ginibre matrices is Gaussian but not identically distributed. The product structure introduces a variance matrix which is averaged (annealed) over a Wishart ensemble. Averaging over Gaussian random couplings is a feature of the Sherrington-Kirkpatrick model \cite{SherringtonKirkpatrick} and more recently the SYK model \cite{K2, Sachdev1993a, Maldacena:2016hyu}, to the best of our knowledge, averaging over the Wishart ensemble is novel.

The connection between RMT and neural networks has a long history; the parameters of a feed forward neural network are neatly packaged into a chain of matrices which certainly hints that RMT could be useful a technique but it is probably fair to say that the current state of the art in this line of research is somewhat short of achieving practical applications with competitive precision. 
We have been particularly influenced by the recent works \cite{Ganguli_exact, Couillet170205419, NIPS2017_6857, 2019arXiv191200827A, yaida2019nongaussian} but drawing a more precise bridge between our current results and neural networks will require further consideration.  Since the number of active degrees of freedom in feed forward neural networks in principle scales greater than $O(N)$, we are particularly interested in results which go beyond the $O(N)$ degrees of freedom one has available in studies of eigenvalues and singular values. So whilst there has been impressive progress in understanding the eigenvalue spectrum of the product of $n$ Ginibre matrices, both at large $N$ \cite{Janik09123422} and finite $N$ \cite{Osborn_2004, Akemann_2012} and also the singular values \cite{PhysRevE.83.061118, Burda_2011, 2013JPhA...46A5205A} we would like to study the mixed moments in the product Gaussian ensemble. 

Perhaps ones first reaction should be somewhat sceptical, these moments have yet to be computed even within the Ginibre ensemble \cite{kemp2009enumeration} and as such it might seem inconceivable that this could then be accomplished in the product ensemble. While this is indeed true, we are optimistic that this problem can be overcome in the Ginibre ensemble and as additional motivation we  ultimately find that the mixed moments in the product ensemble are closely related to the mixed moments in the Ginibre ensemble.

The structure of this paper is as follows: in section \ref{sec:MomentsComplexGaussian} we review some known facts about moments of complex Gaussian matrices, with and without a variance profile. In section \ref{Sec:ProductGinibre} we review some basic and motivational results regarding the eigenvalue and singular value spectrum of the product ensemble. In section \ref{Sec:PDFproduct} we begin our study of the product Ginibre ensemble and derive the probability distribution function as well as our formula for the mixed moments for the case $n=2$. In section \ref{sec:MulitiGinibre} we generalize our results to any positive integer $n$ and present our results for the mixed moments. The main result of our paper is \eq{ProdGinibreMomentFC}. \\

\section{Moments of complex Gaussian matrices}\label{sec:MomentsComplexGaussian}
\subsection{The Ginibre model}\label{Sec:GinibreCorrelators}
The main difficulty encountered in computing expectation values of functions of Ginibre matrices $X$ is the strictly upper triangular term $T$ in the Schur decomposition \eq{SchurDecomp}.  Of course  when $X$ is Hermitian, $T$ vanishes and it follows that in many such cases the computation reduces to an integral over the $N$ eigenvalues $\Lam$. 
The result of Ginibre \cite{Ginibre1965}, in computing the eigenvalue pdf of a complex Gaussian matrix, is an example of a scenario when the integral over the $O(N^2)$ upper triangular terms are all Gaussian and can be performed explicitly. In general however, even at large $N$, moments of Ginibre matrices cannot seemingly be reduced to a problem of $O(N)$ degrees of freedom as the components of $T$ couple non-trivally to the components of $\Lam$. Nonetheless, there are still numerous exactly solvable results in the field of random complex matrices and we now mention a few which we have found to be relevant to the current work.

Expectation values of ratios of products of characteristic polynomials operators\footnote{We use $\langle . \rangle_G$ to denote the expectation value in the Ginibre ensemble.}
\bea
\left\langle  \frac{
\prod_{i=1}^{L_1} \det(\eps_1 - X)
\prod_{i=1}^{L_2}  \det(\eps_2 - X^\dagger) 
}{
\prod_{i=1}^{M_1} \det(\eta_1 - X)
\prod_{i=1}^{M_2} \det(\eta_2 - X^\dagger)
}\right\rangle_G
\eea
have been progressively computed in \cite{Akemann_2001, Akemann_2003, Akemann_2004, bergre2004biorthogonal, Bergere_2006} and the structure of these results follows closely the work done in the Hermitian case \cite{2000CMaPh.214..111B, Fyodorov_2003}. From these results one can derive expectation values for multi-trace operators of the form 
\be
\langle \Tr X^{k_1} \Tr (X^\dagger)^{l_1}\ldots \Tr X^{k_n} \Tr (X^\dagger)^{l_n}\rangle_G
\ee 
but not non-trivial expectation values of single trace operators, since the expectation values of (anti)-holomorphic single trace operators vanish by Wick's theorem:
\bea
\langle \Tr X^k \rangle_G = \langle \Tr (X^\dagger)^k \rangle_G =0\,.
\eea

Indeed, the non-trivial single-trace correlators in the Ginibre ensemble are of the form
\bea
C_{{\bf i}, {\bf j}} &=& \langle\cO_{{\bf i}, {\bf j}} (X, X^\dagger)  \rangle_G \label{GinibreMoments}
\eea
with
\bea
\cO_{{\bf i}, {\bf j}} (X, X^\dagger) &=& \Tr X^{i_1} (X^\dagger)^{j_1}\ldots X^{i_k} (X^\dagger)^{j_k} \label{Odef}
\eea
and 
\bea
{\bf i} &=& (i_1\,,\ldots ,i_k)\,,\quad {\bf j} = (j_1\,,\ldots ,j_k)\\
\sum_{a} i_a &=& \sum_{a} j_a \equiv m \,.
\eea
At large $N$, to leading order, Wick's Theorem reduces the computation of $C_{{\bf i}, {\bf j}} $ to the enumeration of {\it non-crossing pairings} \cite{nica2006lectures} and the state of the art regarding their enumeration can be found in \cite{kemp2009enumeration, Schumacher2013}. The complete enumeration of non-crossing pairings and thus the evaluation of mixed moments in the Ginibre ensemble is currently unsolved, but for example in \cite{kemp2009enumeration} one can find the following explicit result for a subset of mixed moments\footnote{
see also \cite{alexeev2010asymptotic} where this is obtained from the distribution of singular values of $X^s$
}
\bea
\langle  \Tr  \blp X^s(X^\dagger)^s\brp^r\rangle_G = FC_{s}(r)\,.\label{FussCatalanCorrelators}
\eea
which are the Fuss-Catalan numbers (see appendix \ref{app:FussCatalan}). When $s=1$, these moments are equivalent to moments of the Marchenko-Pastur distribution of singular values of $X$ and are equal to the Catalan numbers.

\subsection{Enumerating operators in the Ginibre model}

It is interesting to count the number of operators \eq{Odef} in the Ginibre model with $m$ insertions of both $X$ and $X^\dagger$ and see the growth for large $m$. The counting of operators in the product Ginibre model is identical. Since the trace induces an invariance under cyclic rotation of the $X,X^\dagger$ insertions, this is equivalent to the number of 2-ary necklaces with $m$-beads of each color \cite{OEISA003239} and one can compute this from the cycle index for the cyclic group \cite{TuckerCombinatorics}:
\bea
Z_{C_{2m}} &=& \frac{1}{2m} \sum_{d|2m} \vphi(d) a_d^{2m/d}.\label{CycleIndex}
\eea
where $\vphi(d)$ is Eulers totient function.
We can extract the number of necklaces with $m$ beads of each color from \eq{CycleIndex} by extracting the coefficient of $x^my^m$ from \eq{CycleIndex} with $a_d$ replaced by $x^d + y^d$:
\bea
N_{2m} &=&  \frac{1}{2m} \sum_{d|2m} \vphi(d) (x^d + y^d)^{2m/d} {\Big |}_{x^my^m}  \\
&=&  \frac{1}{2m} \sum_{d|m} \vphi(d) \binom{2m/d}{m/d}\,. \label{N2mPolya}
\eea

The largest contribution to $N_{2m}$ comes from term $d=1$ and this term alone gives the bound
\be
N_{2m} \geq \frac{1}{2m} \binom{2m}{m}
\ee
which for large $m$ approximates the Catalan number $C_m$ and grows as 
\be
C_m \sim \frac{4^m}{\pi^{1/2}m^{3/2}}\,.
\ee
From this counting we quantify the obvious fact that the number of inequivalent operators in the Ginibre model scales much larger than the Hermitian model. We see this as motivation for studying these moments in more detail.

\subsection{Non-trivial variance profile}

An interesting direction of research is to generalize the Ginibre ensemble such that elements in $X$ are no longer i.i.d. but rather there is a non-trivial variance profile\footnote{
Some excellent reviews are \cite{TulinoVerduBook, forrester2010log, BaiSilverteinBook, Couillet:2011:RMM:2161679}.  
}. When this variance profile is set by some fixed, Hermitian positive definite matrix
$W$  we denote these moments by\footnote{
In \eq{CijWdef} the deterministic matrix $W$ appears within the operator insertion itself but using the identity
\bea
  \int d^2X\,  f\blp X W^{1/2}\brp e^{- N \Tr XX^\dagger} &=& \frac{1}{\det W^N}\int d^2X\,  f(X) e^{- N \Tr X W^{-1} X^\dagger}
\eea
we see that up to a normalization, this is equivalent to sampling $X$ from a Gaussian distribution with variance profile given by $W$. 
}
\bea
\tC_{{\bf i}, {\bf j}}(W) &=& \langle\cO_{{\bf i}, {\bf j}} (XW^{1/2}, W^{1/2} X^\dagger)  \rangle_G \label{CijWdef}\\
&=& \frac{1}{N^m}\sum_{k=1}^m  \sum_{\bm \in \cM_{m,k}}
\bslb N^{m-k}\tc_{{\bf i}, {\bf j}} (m_1\,,\ldots ,m_k)\bsrb
 \Tr W^{m_1} \ldots  \Tr W^{m_k}   \label{tCijW}
\eea
for some set of coefficients $\tc_{{\bf i}, {\bf j}}$ and where we have defined the summation
\bea
\sum_{\bm \in \cM_{m,k}}  = \sum_{\substack{m_1+\ldots + m_k=m  \\ m_1\leq m_2\leq \ldots m_k}}\,.
\eea
We have introduced factors of $N$ such that the $\tc_{{\bf i}, {\bf j}}$ are finite at large $N$. We note that \eq{tCijW} defines the coefficients $\tc_{{\bf i}, {\bf j}}$ and is exact in $N$.
In this work we will need one particular result found in \cite{TulinoVerduBook, YIN198650, LiTulinoVerdu} where these authors have considered the expectation value of the operator
\bea
\cO_{1^m,1^m}(X W^{1/2}) \label{O11def}
\eea
and found that to leading order in $\frac{1}{N}$  
\bea
\tc_{1^m,1^m} (m_1\,,\ldots ,m_k) &=& \frac{m!}{(m-k+1)! f(m_1\,,\ldots ,m_k)} \label{XWXcoeffs}
\eea
where
\bea
f(m_1\,,\ldots ,m_k) = \prod_{j=1}^m f_j!
\eea
and $f_j$ is the number of elements of $[m_1\,,\ldots ,m_k]$ which equal $j$. In finite $N$ there will be corrections to $\tc_{1^m,1^m} (m_1\,,\ldots ,m_k)$ from non-planar diagrams.

We note that while in the absence of the variance profile $W$, the order at which a particular ribbon diagram contributes is dictated by its Euler characteristic \cite{'tHooft:1973jz, Brezin:1977sv}, the combinatorics giving rise to \eq{XWXcoeffs} goes beyond enumerating the number of faces, edges and vertices.

\section{Product of Ginibre matrices} \label{Sec:ProductGinibre}

Much is known about the eigenvalue and singular value spectrum of the product of Ginibre matrices and we summarize some of these results here. 

\subsection{Eigenvalues}

Let $\{A_i| i=1,\ldots n\}$ be $N\times N$ Ginibre matrices where the real and imaginary parts of each entry are sampled from $\cN(0, \frac{\sig_i^2}{2N})$ so that 
\be
\langle A_{i,jk} \rangle =0 \,,\qquad  \langle |A_{i,jk}|^2 \rangle = \frac{\sig_i^2}{N}\,.
\ee
or more precisely 
\bea
P(A_i) &=&\left(\frac{N}{\pi \sig_i^2}\right)^{N^2} \int dA_idA_i^\dagger e^{- \frac{N}{\sig_i^2}\Tr A_i A_i^\dagger}\,.\label{GinibreProbDefinition}
\eea
We want to study the matrix product of the $A_i$ 
\bea
X_{(n)} &=& A_1\ldots A_n \,.
\eea

\begin{figure}[!htb]
\begin{subfigure}{.33\textwidth}
  \centering
\captionsetup{justification=centering,margin=1cm}
  \includegraphics[width=.8\linewidth]{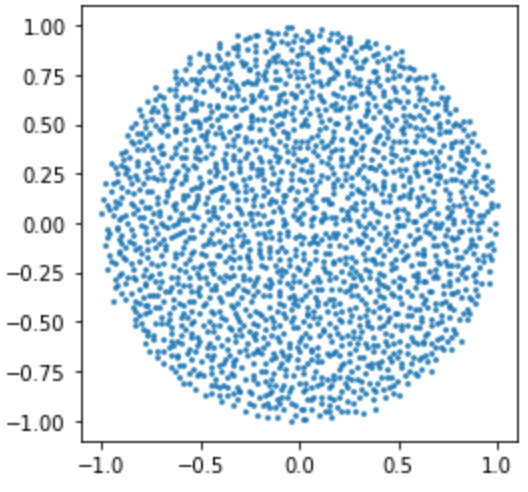}
  \caption{Eigenvalue distribution for $X=A_1$}
\label{Fig:evals1}
\end{subfigure}%
\begin{subfigure}{.33\textwidth}
  \centering
\captionsetup{justification=centering,margin=1cm}
  \includegraphics[width=.8\linewidth]{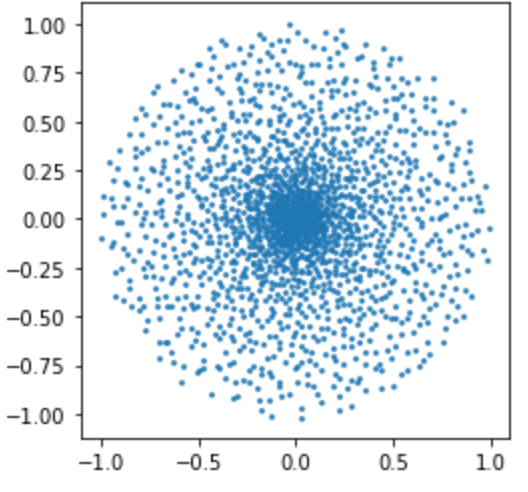}
  \caption{Eigenvalue distribution for $X=A_1A_2$}
\label{Fig:evals12}
\end{subfigure}%
\begin{subfigure}{.33\textwidth}
  \centering
\captionsetup{justification=centering,margin=1cm}
  \includegraphics[width=.8\linewidth]{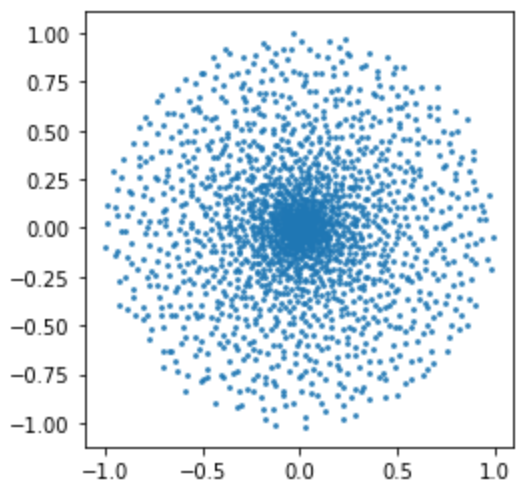}
  \caption{Eigenvalue distribution for $X=A_1A_2A_3$}
\label{Fig:evals123}
\end{subfigure}%
\caption{The distribution of eigenvalues, computed for $N=2000$}
  \label{Fig:Eigenvals}
\end{figure}

The fundamental result in the theory of random complex matrices is due to Ginibre \cite{Ginibre1965} and known as the circular law; to leading order in an expansion in $\frac{1}{N}$, the eigenvalues of a  complex Gaussian $N\times N$ matrix are uniformly distributed on the disk. This has been extended to $X_{(n)}$ in\footnote{See \cite{Burda_2011} for an interesting derivation using free probability and also \cite{gtze2010asymptotic} \cite{ORourkeSoshnikov}} \cite{Janik09123422}. This remarkably simple result for the distribution of eigenvalues of $X_n$ at large $N$ is 
\bea
\rho_n(z,\zbar) &=&\begin{cases}
 \frac{1}{n \pi} \sig^{-\frac{2}{n}} |z|^{-2+\frac{2}{n}} & |z| \leq  \sig\\
 0  & |z|>\sig
 \end{cases} \label{XnEigenvalues}
\eea
and one can see in figure \ref{Fig:Eigenvals} that the for $n>1$, the eigenvalues are no longer uniformly distributed but are more dense near the origin. 

At finite $N$, the eigenvalue pdf of $X_{(n)}$ has been computed in \cite{Akemann_2012} using an inspired matrix factorization of $X_{(n)}$ and the large $N$ limit is shown to agree with \eq{XnEigenvalues}.

\subsection{Singular values}\label{Sec:ProductSingularValues}

The probability density of singular values of $X_{(n)}$ has been computed at large $N$ using a variety of methods  \cite{Banica_2011, benaych-georges2010, Muller1013149, PhysRevE.83.061118, Burda_2010} and at finite in $N$ in \cite{2013JPhA...46A5205A}, once again using an ingenious matrix factorization and orthogonal polynomials. The moments $ \langle (X_{(n)} X_{(n)}^\dagger)^m\rangle $ are integrals of this density of singular values and were in fact computed somewhat earlier in \cite{PhysRevE.83.061118, Banica_2011, benaych-georges2010, Muller1013149}. The upshot for the current work is that the moments are given by Fuss-Catalan numbers (see appendix \ref{app:FussCatalan})
\bea
\frac{1}{N}  \langle (X_{(n)} X_{(n)}^\dagger)^m \rangle &=& FC_{n}(m)\,.\label{SingularMomentsFC}
\eea
Interestingly, comparing \eq{FussCatalanCorrelators} we see the following equality
\bea
\frac{1}{N}  \langle \Tr (X_{(n)} X_{(n)}^\dagger)^m \rangle &=& \frac{1}{N}  \langle \Tr (X^n_{(1)} (X^n_{(1)})^\dagger)^m \rangle \label{MomentEquality1}
\eea
so that these moments are equal if we sample the $A_i$ identically, which is certainly not true of more general mixed moments.

\section{Product of two Ginibre matrices}\label{Sec:PDFproduct}

We will first derive our results for $n=2$, then in section \ref{sec:MulitiGinibre} we will work with $n>2$, where additional features arise. 

To  derive the probability distribution function (pdf) of $X= A_1 A_2$, we use a matrix delta function:
\bea
P_{(2)}(X,X)&= & \left(\frac{N^2}{\pi^2\sig_1^2\sig_2^2}\right)^{N^2} \int \cD A \delta(X -  A_1 A_2) 
e^{-\frac{N}{\sig_1^2}\Tr A_1A_1^\dagger -  \frac{N}{\sig_2^2}\Tr A_2A_2^\dagger}
\eea
where
\be
\cD A = \prod_{i=1}^2 dA_idA_i^{\dagger}\,.
\ee
Using the integral form of $\delta(X)$
\bea
\delta(X) &=& \frac{1}{\pi^{2N^2}} \int dT dT^\dagger e^{i\Tr TX + i \Tr T^\dagger X^\dagger}\,,
\eea
we complete the square for $A_2$ and integrate it out to find
\bea
P_{(2)}(X,X^\dagger)&=&\left(\frac{N}{\pi^3 \sig_2^2}\right)^{N^2}\int d^2T
d^2A_1\, e^{i\Tr TX + i \Tr T^\dagger X^\dagger}  e^{-\frac{N}{\sig_1^2}\Tr A_1 A_1^\dagger(1\!\!1 + \frac{\sig^2}{N^2}T^\dagger T)} \\
&=&\frac{1}{\pi^{2N^2}} \int dTdT^\dagger \frac{e^{i\Tr TX + i \Tr T^\dagger X^\dagger}} {\det(1+ \frac{\sig^2}{N^2} T^\dagger T)^{N}} \label{PXdet}\,.
\eea
where we have defined 
\be
\sig = \sig_1\sig_2.
\ee
We define the moment generating function to be
\bea
\vphi_{(2)}(T,T^\dagger) &=&\pi^{2N^2}  \int dTdT^\dagger e^{-i\Tr TX - i \Tr T^\dagger X^\dagger} P(X,X^\dagger)
\eea
so we can generate the moments by derivatives on $\vphi(T,T^\dagger)$
\bea
\langle X_{i_1\ibar_1} \ldots X_{i_m\ibar_m} X^\dagger_{\jbar_1 j_1} X^\dagger_{\jbar_n j_n}  \rangle 
&=& \left.(-i)^{m+n} \del_{T_{i_1\ibar_1}}\ldots \del_{T_{i_m\ibar_m}} \del_{T^\dagger_{\jbar_1 j_1}}\ldots \del_{T^\dagger_{\jbar_n j_n}} \phi(T,T^\dagger)\right|_{T=T^\dagger=0}  \,.
\eea
From \eq{PXdet} we have an explicit expression
\bea
\vphi_{(2)}(T,T^\dagger) 
&=&\frac{N^{N^2}}{\Gam_N(N)}  \int_{W>0} dW\, e^{ - N \Tr W} e^{-\frac{\sig^2}{N}\Tr T^\dagger T W}\,,  \label{MGFinW}
\eea
where we are using 
\be\label{eq: schwinger determinant}
\frac{1}{\det \Sig^{a}} = \frac{1}{\Gam_N(a)} \int_{W>0} dW \det W^{a-N} e^{-\Tr \Sig\, W}\,.
\ee
The expression \eq{eq: schwinger determinant} gives the determinant of an $N\times N$ 
positive definite Hermitian 
matrix in terms of the complex multivariable gamma function $\Gam_N(a)$ (see appendix \ref{app:MultiGamma} for details) and might be thought of as a multi-variable Schwinger trick. The $W$ appearing in Equation 
\eqref{eq: schwinger determinant} is scaled by $N$ to obtain \eqref{MGFinW},
and the integration is over the space of positive definite Hermitian matrices. 
We can then perform the Fourier transform of the moment generating function and integrate out $T$ to obtain an integral expression for the pdf of $X$:
\bea
P_{(2)}(X,X^\dagger)&=&\frac{N^{N^2}}{\Gam_N(N)} \left(\frac{N}{\pi\sig^2}\right)^{N^2} \int_{W>0} dW\, \frac{e^{ - N\,\Tr W} }{\det W^N}e^{-\frac{N}{\sig^2}\Tr XX^\dagger W^{-1}} \,. \label{PDFX2}
\eea
It is straightforward to verify that
this pdf is normalized to 1.

\subsection{Moments in the product Ginibre model}

Our expression for the pdf \eq{PDFX2} might be recognized as a matrix Bessel function  but we will find it more effective to view it as Gaussian in $X,X^\dagger$ with a variance matrix which is averaged over a Wishart measure. So we see that, as mentioned in the introduction, the product of two Ginibre matrices remains Gaussian but is no longer identically distributed.

We can compute the mixed moments of $X_{(2)}$
\bea
C^{(2)}_{{\bf i,j}} &=& \langle  \cO_{{\bf i,j}}(X, X^\dagger) \rangle_{(2)} \\
&=& \frac{N^{N^2}}{\Gam_N(N)} \left(\frac{N}{\pi\sig^2}\right)^{N^2} \int_{W>0} dW\, \frac{e^{ - N\,\Tr W} }{\det W^N}
\int d^2X\, \cO_{{\bf i,j}}(X, X^\dagger) \,\,e^{-\frac{N}{\sig^2}\Tr X X^\dagger W^{-1}}  
\eea
in two steps:
\begin{enumerate}
\item  Compute the corresponding moment $\tC_{{\bf i,j}}(W)$ in the Ginibre model
\bea
\tC_{{\bf i,j}}(W) &=&  \left(\frac{N}{\pi\sig^2}\right)^{N^2}
\frac{1}{\det W^N}
\int d^2X\, \cO_{{\bf i,j}}(X,X^\dagger) \,\,e^{-\frac{N}{\sig^2}\Tr X X^\dagger W^{-1}} \label{eq:expectationGinibreW}
\eea
in terms of products of traces of $W$.
\item Average the resulting traces against  the Wishart measure\footnote{There are more general Wishart measures which would arise from the product of rectangular Gaussian matrices but we will not consider them here.}
\be
P_W(W) = \frac{N^{N^2}}{\Gam_N(N)}\, e^{ - N \Tr W}\,. \label{WLProbDefintiion}
 \ee
\end{enumerate}
It follows from Wick's theorem that there are $m=\sum_{a} i_a=\sum_{a} j_a$ insertions of $W$ but determining the trace structure in \eq{tCijW} by evaluating the coefficients $\tc_{{\bf i}, {\bf j}}$  is harder than computing $C_{{\bf i}, {\bf j}}$ which is already unsolved in the general case. The result from the first step is of the form
\be \label{CtildeWTraces}
\tC_{{\bf i,j}}(W)  =  \frac{\sig^{2m}}{N^m}  \sum_{k =1}^m \sum_{\bm \in \cM_{m,k}}  \bslb N^{m-k} \tc_{{\bf i,j}}(m_1\,,\ldots , m_k)\bsrb \Tr W^{m_1} \ldots \Tr W^{m_k}\,.
\ee
The second step of averaging over the Wishart distribution is,
to leading order in $N$, quite straightforward. The multi-trace expectation value factorizes into a product of single trace expectation values 
\be
\frac{1}{N^k}\langle\Tr W^{m_1} \ldots \Tr W^{m_k} \rangle_W = \frac{1}{N^k} \langle\Tr W^{m_1} \rangle_W \ldots \langle\Tr W^{m_k} \rangle_W
\ee
and
\bea
\frac{1}{N} \langle \Tr W^l \rangle =C_l
\eea
where $C_l$ are the Catalan numbers. This enables us to write the following
expression, valid in the large $N$ limit:
\be
\boxed{C^{(2)}_{{\bf i,j}} =  \sig^{2m}\sum_{k =1}^m \sum_{\bm \in \cM_{m,k}} \tc_{{\bf i,j}}(m_1\,,\ldots , m_k) C_{m_1}\ldots C_{m_k}  }\label{X2momentResult}\,,
\ee
where $\tc_{\bf i,j}(m_1\,,\ldots , m_k)$ should be computed by enumerating particular planar diagrams. As remarked previously, this enumeration has been carried out in
the literature for certain operators $\mathcal{O}\left(X,X^\dagger\right)$.

\subsection{Examples}
Here  we compute by hand some mixed moments of $X_{(2)}, X_{(2)}^\dagger$ at leading order in  $\frac{1}{N}$ using our prescription as well as some examples of the computation of \eq{CtildeWTraces} at finite $N$ \ref{sec:FiniteN}. We use Wick's theorem with the propagator
\bea
\wick{\c1 X_{ij} \c1 X_{kl}^\dagger} &=& \frac{\sig^2}{N}\delta_{il} W_{jk}
\eea
to evaluate the expectation value \eq{eq:expectationGinibreW} in the Ginibre ensemble. 
In this way for example one finds
\bea
\tC_{{\bf 1,1}}(W) = \frac{1}{N} \wick{\c1 X_{ij} \c1X^\dagger_{ji}} \,
= \frac{ \sig^2}{N} \Tr W \label{2ptfunction}
\eea

\subsubsection{$\frac{1}{N}\langle  \Tr (X_{(2)}X_{(2)}^\dagger )^m\rangle$}\label{sec:Simplen2Examples}

As mentioned in section \ref{Sec:ProductSingularValues}, expectation values of $\langle \Tr  (X_{(2)}X_{(2)}^\dagger )^m\rangle$ are already known from the study of singular values of $X_{(n)}$. Here we will re-compute them as an exposition and check of our formalism.

\vskip 5mm
\noindent ${\bf m=1}$ \\
From \eq{CtildeWTraces} we find 
\bea
\frac{1}{\sig^2}\tC_{1,1}(W) &=&  \frac{1}{N} \tc_{1,1}(1) \Tr W
\eea
with 
\be
\tc_{1,1}(1) =1
\ee 
and thus $\tC_{1,1}(W)=\sig^2 \Tr W$ and trivially agrees with \eq{2ptfunction}. We then have 
\bea
\frac{1}{\sig^2}C^{(2)}_{1,1} &=& \tc_{1,1}(1) C_1 \\
&=&1\,.
\eea
This agrees with \eq{SingularMomentsFC} since $FC_2(1)=1$.

\begin{figure}[!htb]
\begin{subfigure}{.5\textwidth}
  \centering
\captionsetup{justification=centering,margin=1cm}
  \includegraphics[width=.4\linewidth]{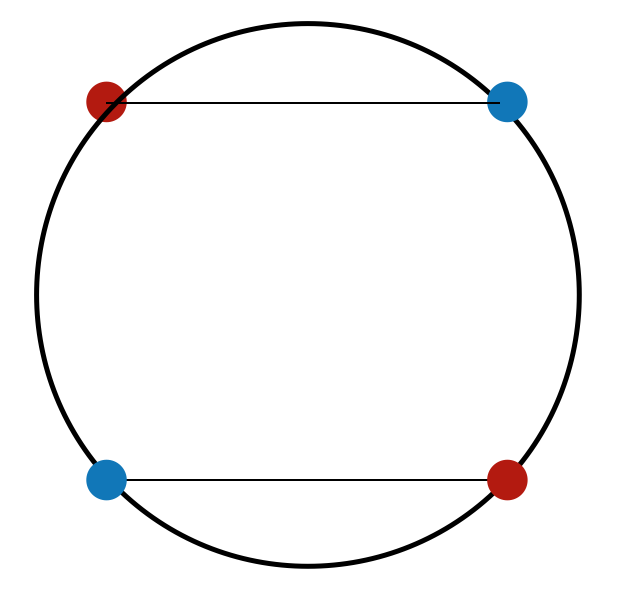}
\caption{Pairing which gives $\Tr W^2$}
\label{Fig:X4diag1}
\end{subfigure}%
\begin{subfigure}{.5\textwidth}
  \centering
\captionsetup{justification=centering,margin=1cm}
  \includegraphics[width=.4\linewidth]{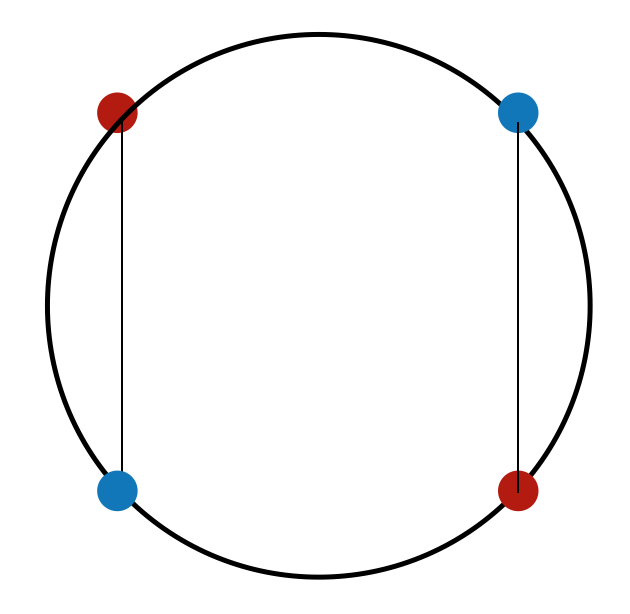}
\caption{Pairing which gives $(\Tr W)^2$}
\label{Fig:X4diag2}
\end{subfigure}%
\caption{Pairings for $\langle X_{(2)} X_{(2)}^\dagger X_{(2)}  X_{(2)}^\dagger \rangle$, both are non-crossing}
  \label{Fig:X4diags}
\end{figure}

\vskip 5mm
\noindent ${\bf m=2}$ \\
From \eq{CtildeWTraces} we find
\bea
\frac{1}{\sig^4}\tC_{1^2,1^2}(W) &=&  \frac{1}{N}  \tc_{1^2,1^2}(2) \Tr W^2  +  \frac{1}{N^2}  \tc_{1^2,1^2}(1,1)( \Tr W )^2 \label{tC2}
\eea
where from \eq{XWXcoeffs} we have
\bea
\tc_{1^2, 1^2}(2) &=& 1 \,,\quad \tc_{1^2, 1^2}(1,1) = 1\,.
\eea
It might be instructive to be painfully explicit and demonstrate how \eq{tC2} arises from non-crossing pairings, the two diagrams are in figures \ref{Fig:X4diags}. The Ginibre moment with explicit indices is
\bea
\frac{1}{\sig^4} \tC_{1^2,1^2}(W) &=& \frac{1}{N} \langle X_{ij}X_{jk}^\dagger X_{kl}X_{li}^\dagger \rangle 
\eea
which by Wick's theorem gives
\bea
\frac{1}{\sig^4} \tC_{1^2,1^2}(W) &=& \frac{1}{N^3} \delta_{ik}W_{jj}    \delta_{ki}W_{ll} + \frac{1}{N^3}   \delta_{ii}W_{jl}    \delta_{kk}W_{lj}  \\
&=&\frac{1}{N^2} (\Tr W)^2  + \frac{1}{N} \Tr W^2 \label{tC4}
\eea
With red dots representing $X$ and blue dots representing $X^\dagger$, the diagram  \ref{Fig:X4diag1} gives the first term in \eq{tC4} and \ref{Fig:X4diag2} gives the second. This example is slightly too simple in that all pairings are non-crossing but this is most certainly not true in the general case. We then obtain $C^{(2)}_{1^2,1^2}$ by averaging over the Wishart ensemble
\bea
\frac{1}{\sig^4} C^{(2)}_{1^2,1^2} &=& \frac{1}{N^2} \langle \Tr W \rangle^2 + \frac{1}{N} \langle \Tr W^2\rangle \\
&=& C_1^2 + C_2 \\
&=& 3
\eea 
which agrees with \eq{SingularMomentsFC} since $FC_2(2)=3$.

\begin{figure}[!htb!]
\begin{subfigure}{.5\textwidth}
  \centering
\captionsetup{justification=centering,margin=1cm}
  \includegraphics[width=.4\linewidth]{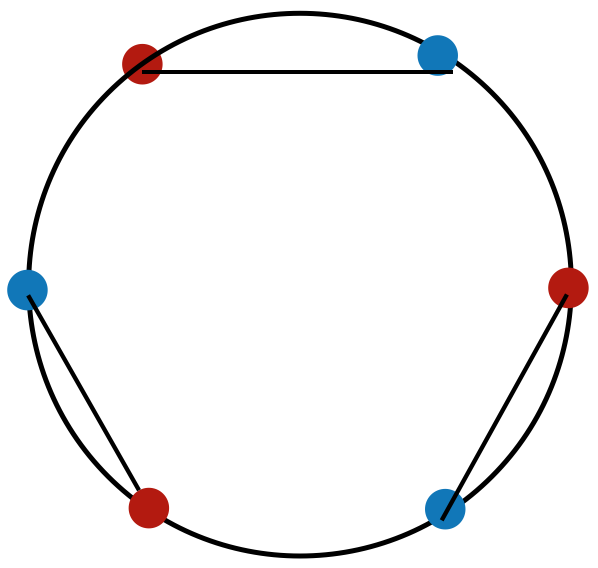}
\caption{Pairing which gives $\frac{1}{N^3}(\Tr W)^3$}
\label{Fig:necklace6_1}
\end{subfigure}%
\begin{subfigure}{.5\textwidth}
  \centering
\captionsetup{justification=centering,margin=1cm}
  \includegraphics[width=.4\linewidth]{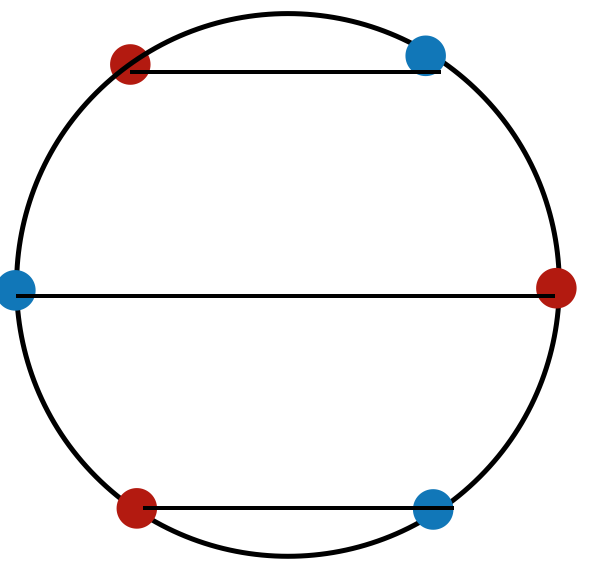}
\caption{Pairing which gives $\frac{1}{N^2}\Tr W \Tr W^2$}
\label{Fig:necklace6_2}
\end{subfigure}%
\\
\begin{subfigure}{.5\textwidth}
  \centering
\captionsetup{justification=centering,margin=1cm}
  \includegraphics[width=.4\linewidth]{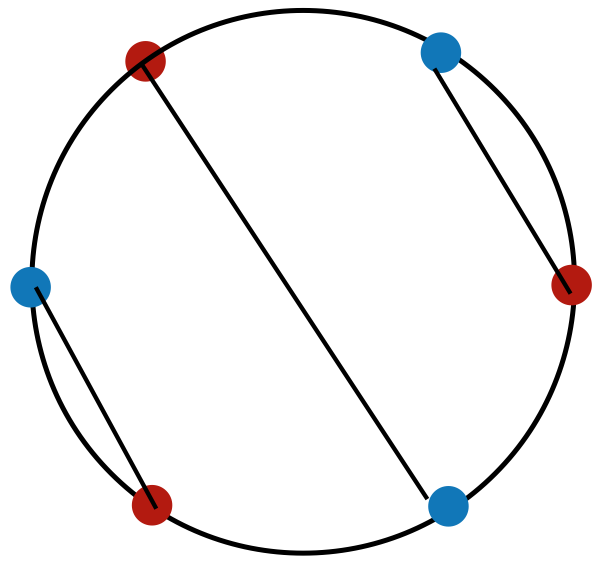}
\caption{Pairing which gives $\frac{1}{N^2}\Tr W \Tr W^2$}
\label{Fig:necklace6_3}
\end{subfigure}%
\begin{subfigure}{.5\textwidth}
  \centering
\captionsetup{justification=centering,margin=1cm}
  \includegraphics[width=.4\linewidth]{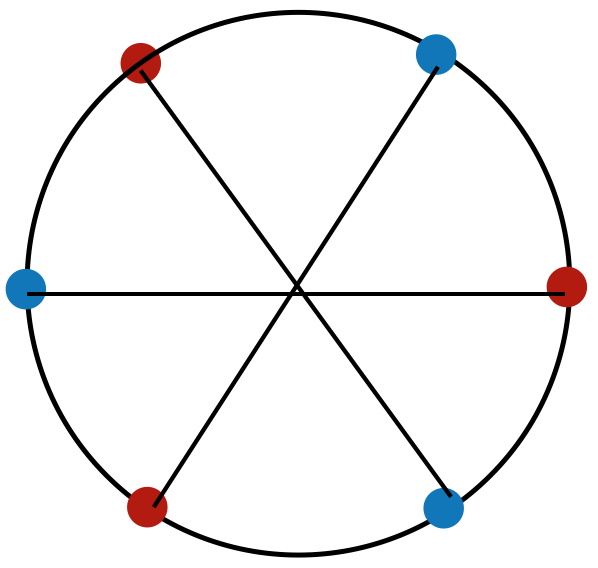}
\caption{Pairing which gives the subleading term $\frac{1}{N^3}\Tr W^3$}
\label{Fig:necklace6_4}
\end{subfigure}%
\\
\begin{subfigure}{.5\textwidth}
  \centering
\captionsetup{justification=centering,margin=1cm}
  \includegraphics[width=.4\linewidth]{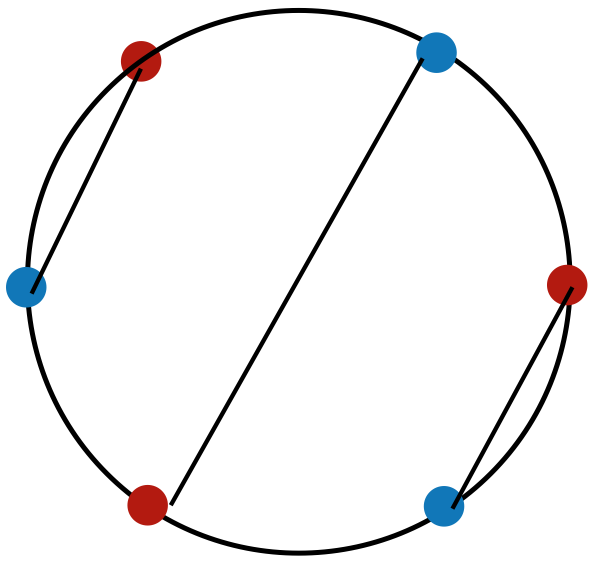}
\caption{Pairing which gives $\frac{1}{N^2}\Tr W \Tr W^2$}
\label{Fig:necklace6_5}
\end{subfigure}%
\begin{subfigure}{.5\textwidth}
  \centering
\captionsetup{justification=centering,margin=1cm}
  \includegraphics[width=.4\linewidth]{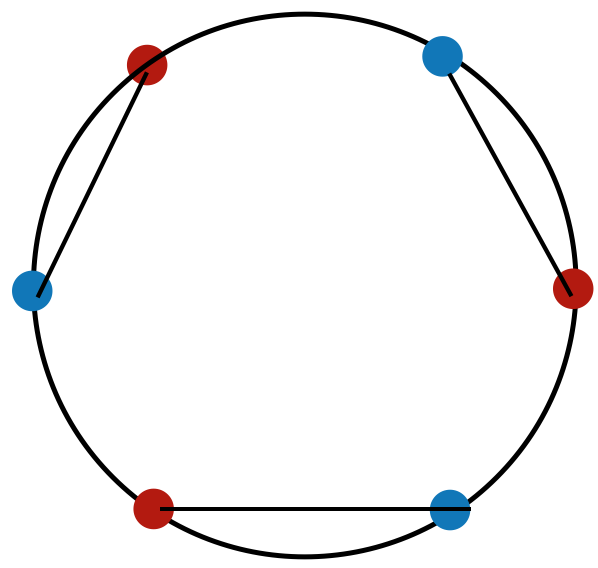}
\caption{Pairing which gives $\frac{1}{N}\Tr W^3$}
\label{Fig:necklace6_6}
\end{subfigure}%
\caption{Pairings which arise from $\frac{1}{N} \langle \Tr (X X^\dagger)^3  \rangle$. Diagram \ref{Fig:necklace6_4} is crossing and thus contributes only at subleading order.}
\label{Fig:necklaces6}
\end{figure}

\vskip 5mm
\noindent ${\bf m=3}$ \\
We will present one more example in full detail as it exhibits some richer features than $m=2$ above. The Ginibre moment is given by
\bea
\frac{1}{\sig^6} \tC_{1^3,1^3}(W) &=&\frac{1}{N} \langle \Tr X X^\dagger X X^\dagger X X^\dagger \rangle \\
&=&  \frac{1}{N}\langle X_{ij} X_{jk}^\dagger X_{kl} X_{lm}^\dagger X_{mn} X_{ni}^\dagger \rangle \\
&=& \frac{1}{N^4}W_{jj} \delta_{ik} (W_{ll} \delta_{km} W_{nn} \delta_{mi} +W_{ln} \delta_{ki} W_{nl} \delta_{mm}  )  \non \\
&& +\frac{1}{N^4} W_{jl} \delta_{im} (W_{lj} \delta_{kk} W_{nn} \delta_{mi} +W_{ln} \delta_{ki} W_{nj} \delta_{mk}  )  \non \\
&& +\frac{1}{N^4} W_{jn} \delta_{ii} (W_{ll} \delta_{km} W_{nj} \delta_{mk} +W_{lj} \delta_{kk} W_{nl} \delta_{mm}  )  \\
&=& \frac{1}{N^3} (\Tr W)^3 +  \frac{1}{N^2} \Tr W \Tr W^2 \non \\
&&  +  \frac{1}{N^2} \Tr W \Tr W^2  + \frac{1}{N^3} \Tr W^3  \non\\
&&  + \frac{1}{N^2}\Tr W  \Tr W^2  +  \frac{1}{N}  \Tr W^3  \label{XXdag3}
\eea 
In figures \ref{Fig:necklaces6} we have the pairings which give rise to \eq{XXdag3}, hopefully it is clear from their arrangement which diagram corresponds to which contractions. We find that the trace structure which weights each diagram is not obvious from the diagram itself, in particular it is a finer structure than enumerating vertices, edges and faces on its corresponding fatgraph.

We have from \eq{CtildeWTraces}
\bea
\frac{1}{\sig^6} C^{(2)}_{1^3,1^3} &=&  \frac{1}{N}\tc_{1^3,1^3}(3) \Tr W^3  
+  \frac{1}{N^2}\tc_{1^3,1^3}(1,2)\Tr W \Tr W^2 
+  \frac{1}{N^3} \tc_{1^3,1^3}(1,1,1)(\Tr W)^3
\eea
and obtain the large $N$ results from \eq{XWXcoeffs}
\bea
\tc_{1^3,1^3}(1,1,1) &=& 1\,,\quad \tc_{1^3,1^3}(1,2)=3\,,\quad \tc_{1^3,1^3}(3)=1\,.
\eea
We can see from \eq{XXdag3} that $\tc_{1^3,1^3}(1,1,1)$ counts the number of diagrams which give $(\Tr W)^3$, $\tc_{1^3,1^3}(1,2)$ counts the diagrams which give $\Tr W \Tr W^2$ and $\tc_{1^3,1^3}(3)$ counts the diagrams which give $\Tr W^3$.
We finally have
\bea\label{eq: planar corr m eq 3}
\frac{1}{\sig^6} C^{(2)}_{1^2,1^2} &=& C_3 + 3 C_2 +C_1 \\
&=& 12
\eea
which agrees with \eq{SingularMomentsFC} since $FC_2(3)=12$.

\subsubsection{$\frac{1}{N} \langle \Tr X_{(2)}^m (X_{(2)}^\dagger )^m\rangle$}
\label{sec:Example2}

\begin{figure}[!htb]
  \centering
  \includegraphics[width=.2\linewidth]{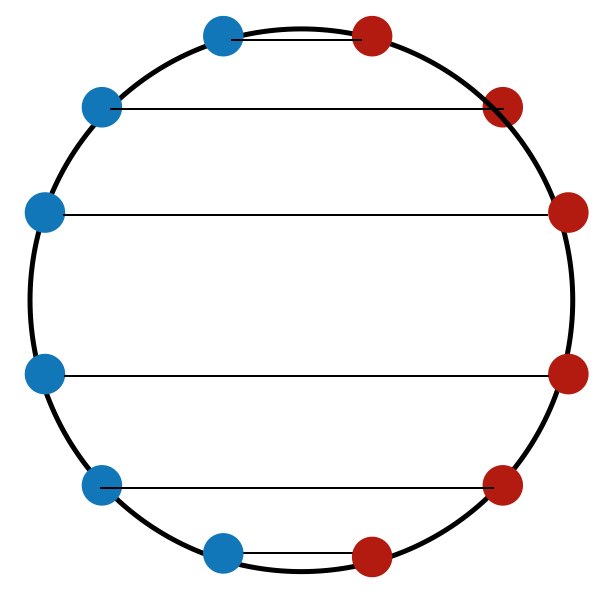}
  \caption{The sole non-crossing pairing for $\langle X_{(2)}^6 (X_{(2)}^\dagger )^6\rangle $}
\label{Fig:necklace1}
\end{figure}

This family of moments is manageable by hand since there is a single non-crossing pairing (see figure \ref{Fig:necklace1}) and this evaluates to 
\bea
\frac{1}{\sig^{2m}}C^{(2)}_{m,m}&=&\frac{1}{N} \langle \Tr X_{(2)}^m (X_{(2)}^\dagger )^m\rangle \\
&=& \frac{1}{N^m}\langle \Tr W \rangle^m \\
&=& 1 
\eea

\subsubsection{$\frac{1}{N}\langle  \Tr X_{(2)}^m X_{(2)}^\dagger X_{(2)} (X_{(2)}^\dagger )^m\rangle$}
\label{sec:Example3}

\begin{figure}[!htb]
\begin{subfigure}{.5\textwidth}
  \centering
\captionsetup{justification=centering,margin=1cm}
  \includegraphics[width=.4\linewidth]{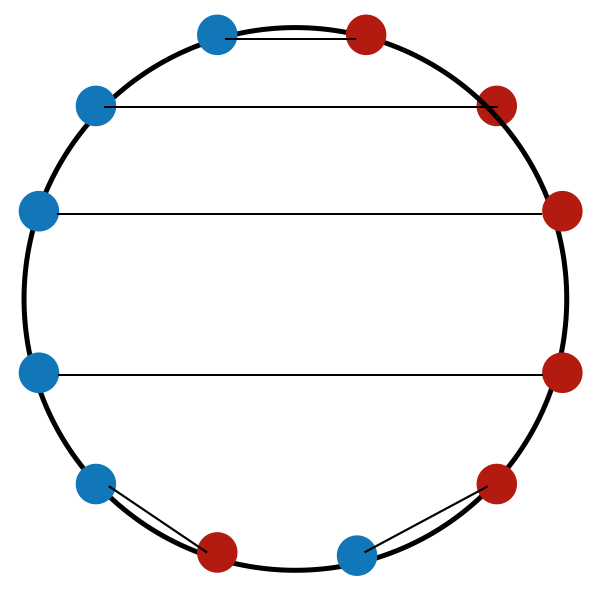}
\label{Fig:necklace2}
\end{subfigure}%
\begin{subfigure}{.5\textwidth}
  \centering
\captionsetup{justification=centering,margin=1cm}
  \includegraphics[width=.4\linewidth]{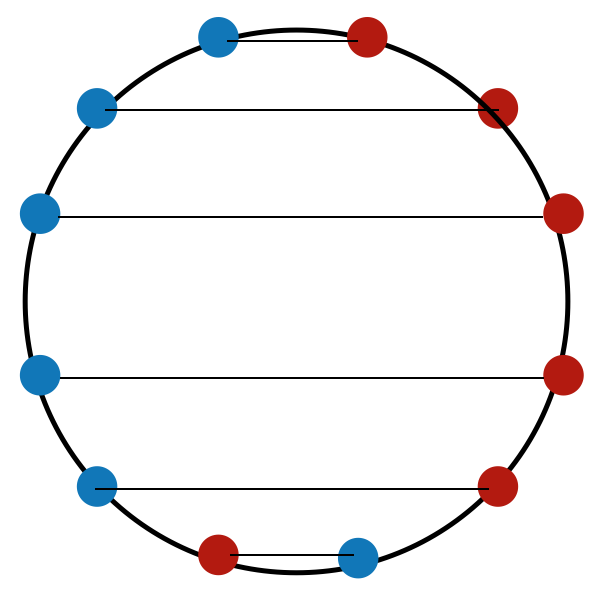}
\label{Fig:necklace3}
\end{subfigure}%
\caption{Non-crossing pairings for $\langle X_{(2)}^5 X_{(2)}^\dagger X_{(2)} (X_{(2)}^\dagger )^5\rangle$}
  \label{Fig:necklaces23}
\end{figure}

This family of moments is also manageable by hand since there are just two non-crossing pairings (see figures \ref{Fig:necklaces23}) and we find that this evaluates to 

\bea
\frac{1}{\sig^{2m}}C^{(2)}_{[m-1,1],[1,m-1]}&=& \frac{1}{N} \langle \Tr X_{(2)}^m X_{(2)}^\dagger X_{(2)} (X_{(2)}^\dagger )^m\rangle \\
&=&  \frac{1}{N^{m+1}}  \langle \Tr W \rangle^{m+1} + \frac{1}{N^{m}} \langle \Tr W^2 \rangle\langle \Tr W \rangle^{m-1}  \\
&=& 3\,.
\eea
Our numerical experiments are summarized in table \ref{table:moment2} and give excellent agreement. 
\begin{table}[bth!]
      \centering
\[\begin{array}{c|ccccc}
 m & \frac{1}{N}\langle \Tr X_{(2)}^m X_{(2)}^\dagger X_{(2)} (X_{(2)}^\dagger )^m\rangle\\
\hline
1 & 2.998 \\
2 &2.998 \\
3 &2.998 \\
4 &2.997 \\
\end{array}\]
\caption{Numerical values of $\frac{1}{N} \langle \Tr X_{(2)}^m X_{(2)}^\dagger X_{(2)} (X_{(2)}^\dagger )^m\rangle$ with $N=500$ and averaged from $1500$ samples.}
    \label{table:moment2}
\end{table}

\subsection{Computing moments at finite $N$}\label{sec:FiniteN}
While we have mostly focussed computing the moments in the product
model at leading order in $N$ so far, 
there is no conceptual
difficulty in implementing our prescription
at finite $N$ on a case-by-case basis.

As an illustration, we shall compute the expectation value of 
$\Tr\left(X_{(2)}X_{(2)}^\dagger\right)^3$
without invoking the large $N$ approximation. The computation in the Ginibre
ensemble by enumerating all pairs, both non--crossing and crossing,
has already been done in \eqref{XXdag3}.
We are therefore left
with computing the expectation values in the Wishart ensemble
i.e. step 2 in our prescription. 
A straightforward way of computing these is from the integral
\be\label{eq:int WL}
\mathcal{I}\left(N,T\right) =\frac{N^2}{\Gamma_N\left(N\right)}
\int_{W>0}\, dW\, e^{- \Tr N\, W\,T}
=\frac{1}{\left\vert T\right\vert^{N}}\,,
\ee
in terms of which
\be \label{eq:WL corr}
\left\langle\, W_{i_1j_1}W_{i_2j_2}\,\ldots\,W_{i_mj_m}\,\right\rangle_W
=
\left(-\frac{1}{N}\right)^m \partial_{T_{j_1i_1}}\partial_{T_{j_2i_2}}
\,\ldots\,\partial_{T_{j_mi_m}}
\frac{1}{\left\vert T\right\vert^{N}}\Bigg\rvert_{T_{ij}=\delta_{ij}}\,.
\ee
For the moment of interest, it suffices to compute terms up to $m=3$. We find
\be
\begin{split}
&\left\langle W_{ij}\right\rangle= \delta_{ij}\,,
\qquad
\left\langle W_{ij}W_{kl}\right\rangle= \delta_{ij}\delta_{kl}
+\frac{1}{N}\,\delta_{il}\delta_{kj}\,,\\
\left\langle W_{ij}W_{kl}W_{pq}\right\rangle
= &\delta_{pq}\delta_{kl}\delta_{ij}
+\frac{1}{N}\left(\delta_{pq}\delta_{il}\delta_{kj}
+\delta_{kq}\delta_{pl}\delta_{ij}
+\delta_{kl}\delta_{iq}\delta_{pj}\right)
+\frac{1}{N^2}\left( \delta_{iq}\delta_{pl}\delta_{kj}
+\delta_{il}\delta_{kq}\delta_{pj}\right)\,.
\end{split}
\ee
Hence
\be
\frac{1}{N}\left\langle\Tr W\right\rangle =1\,,
\quad
\frac{1}{N}\left\langle\Tr W^2\right\rangle
=2\,,
\quad \frac{1}{N}\left\langle\Tr W^3\right\rangle
= 5+\frac{1}{N^2}\,,
\ee
and
\be 
\frac{1}{N^2}\left\langle\Tr W^2\Tr W\right\rangle = 2+\frac{2}{N}+\frac{2}{N^2}\,.
\ee
We then use these expectation values with Equation \eqref{XXdag3}
to find that the moment in the product ensemble is given by
\bea
\frac{1}{\sig^6}\,C_{1^3,1^3}
&=& 1+  3 \left(2+\frac{2}{N}+\frac{2}{N^2}\right)
 + \frac{1}{N^2}\left(5+\frac{1}{N^2}\right)  + 
 \left(5+\frac{1}{N^2}\right) \non\\
&=& 12+\frac{6}{N}+\frac{12}{N^2}+\frac{1}{N^4} 
\simeq  12+\frac{6}{N}\,.
\eea
The $O(1)$ term is 12, the result previously obtained in the large
$N$ limit in Equation \eqref{eq: planar corr m eq 3}.

To summarize the corrections to the large $N$ formula in this case,
\begin{enumerate}
\item The contribution of the crossing pair was suppressed by $\frac{1}{N^2}$
with respect to the non-crossing pairs in the Ginibre moment, 
\item The expectation values $\frac{1}{N^k}\Tr\left\langle W^k\right\rangle$ were
given by $C_k+\mathcal{O}\left(\frac{1}{N^2}\right)$, and,
\item We found that 
\be 
\left\langle\Tr W^2\Tr W\right\rangle= 
\left\langle\Tr W^2\right\rangle\left\langle
\Tr W\right\rangle+\mathcal{O}\left(\frac{1}{N}\right)\,,
\ee
i.e. there is a non-trivial $O\left(\frac{1}{N}\right)$ correction to 
large $N$ factorization in the Wishart ensemble.
\end{enumerate}
From this it appears that the first subleading--in--$N$ correction comes from
the fact that expectation values of multi-trace operators in $W$ no longer 
factorize into products of expectation values of single trace operators.

\section{Product of multiple Ginibre matrices}\label{sec:MulitiGinibre}
In this section we generalize our results from section \ref{Sec:PDFproduct} to the product of  $n$ Ginibre matrices
\be
X_{(n)} = A_1A_2\ldots A_n\,.
\ee
We will again be able to reduce the computation of the mixed moments (denoted by $C^{(n)}_{\bf i,j}$ below) to a computation of the mixed moments in the Ginibre model with deterministic variance profile $\tC_{\bf i,j}(W)$ but whereas the Catalan numbers played a central role in \eq{X2momentResult}, they are now replaced with the Fuss-Catalan numbers defined in Equation \eq{FC_definition}.

The probablility distribution function of the $X_{(n)}$ is given by
\bea
P_{(n)}(X,X^\dagger)&= & \left(\frac{N^{n}}{\pi^n \sig}\right)^{N^{2}} \int \cD^{(n)} A\, \delta(X -  \prod_{i=1}^n A_1) 
e^{- \sum_{i=1}^n \frac{N}{\sig_i^2}\Tr A_iA_i^\dagger}\,,
\eea
where
\bea
\sig &=& \sig_1 \ldots \sig_n\,,\quad  \cD^{(n)}A= \prod_{i=1}^n dA_i dA_i^\dagger\,.
\eea
After some algebra (see appendix \ref{App:MomGenMultiple}), we find
that this reduces to
\bea
P_{(n)}(X,X^\dagger)&= & 
 \frac{N^{(n-1)N^2}}{\Gam_N(N)^{n-1}}
\Blp\frac{N}{\pi \sig^2}\Brp^{N^2}
\int_{W_i>0} \prod_{i=1}^{n-1}\Bslb dW_i \frac{1}{\det W_i^N } e^{-N\Tr W_i} \Bsrb e^{ -\frac{N}{\sig^2}\Tr  W_{(n-1)}^{-1}X X^{\dagger}}\,,\label{eq:ProbGinibre}
\eea
where
\be
W_{(n-1)} = W_1\ldots W_{n-1}
\ee
and the integral is over positive definite Hermitian $W_i$.
It is again straightforward to check that the pdf \eq{eq:ProbGinibre} is unit normalized.

\subsection{Mixed moments at large $N$}

To obtain an expression for the moments we will work to leading order in $\frac{1}{N}$.  The mixed moments in this ensemble are then given by \\
\bea
C^{(n)}_{\bf i,j}&= & \langle\cO_{\bf i,j}(X,X^\dagger)  \rangle_{(n)} \\
&=&   \frac{N^{(n-1)N^2}}{\Gam_N(N)^{n-1}}
\Blp\frac{N}{\pi \sig^2}\Brp^{N^2}  \int_{W_i>0} \prod_{i=1}^{n-1}\Bslb dW_i \frac{1}{\det W_i^N } e^{-\Tr W_i} \Bsrb \int d^2X\, \cO_{\bf i,j}(X,X^\dagger)  e^{ -\frac{N}{\sig^2}\Tr W_{(n-1)}^{-1}X X^{\dagger}}\non \\
&& \label{Cnij1}
\eea
where $\langle\cdot \rangle_{(n)}$ denotes the expectation value in the product Ginibre ensemble \eq{eq:ProbGinibre}.
Using \eq{tCijW} we can obtain from \eq{Cnij1}
\bea
C^{(n)}_{\bf i,j} &=& \langle  \tC_{{\bf i,j}}(W_{(n-1)}) \rangle_{W,n-1} \\
&=&\frac{\sig^{2m}}{N^m}\sum_{k=1}^m \sum_{\bm \in \cM_{m,k}}  
\bslb N^{m-k}\tc_{{\bf i}, {\bf j}} (m_1\,,\ldots ,m_k)\bsrb
\langle \Tr W_{(n-1)}^{m_1} \rangle_{W,n-1} \ldots  \langle\Tr W_{(n-1)}^{m_k} \rangle_{W,n-1} \label{ProdGinibreMoment}
\eea
where $\langle\cdot \rangle_{W,n-1}$ denotes the expectation value in the multi-Wishart ensemble, whose pdf is given by
\bea
P_{W,n-1} &=& \frac{N^{(n-1)N^2}}{\Gam_N(N)^{n-1}}  \prod_{i=1}^{n-1} e^{-N\Tr W_i} \label{eq:WishartnPDF}\,.
\eea
 
To proceed, we must for all $m>0$, compute the mixed moment in the multi-Wishart ensemble to leading order in $\frac{1}{N}$: 
 \bea
\frac{1}{N} \langle \Tr W_{(n-1)}^{m} \rangle_{W,n-1} &=&\frac{N^{(n-1)N^2}}{\Gam_N(N)^{n-1}} \int_{W_i>0} \Bslb \prod_{i=1}^{n-1} dW_i e^{-N\Tr W_i} \Bsrb \frac{1}{N}\Tr (W_1\ldots W_{n-1})^m\,.
\eea
 which for $n=2$ is given by the Catalan numbers. This is where the result  \eq{tCijW} and \eq{XWXcoeffs} is crucial; from the form of the operator insertion, we see that we can solve this moment recursively by first treating $W_1\ldots W_{n-2}=W_{(n-2)}$ as a variance matrix for $X_{n-1}$ where 
 \be
 W_{n-1} = X_{n-1}X_{n-1}^\dagger\,.
 \ee 
 So we have
 \bea
 \frac{1}{N} \langle \Tr W_{(n-1)}^{m} \rangle_{W,n-1} &=&\frac{N^{(n-1)N^2}}{\Gam_N(N)^{n-1}} \int_{W_i>0} \Bslb \prod_{i=1}^{n-2} dW_i e^{-N\Tr W_i} \Bsrb \non \\ 
 &&  \times \Blp \frac{N}{\pi}\Brp^{N^2}\int d^2 X_{n-1} e^{-N\Tr X_{n-1}X_{n-1}^\dagger}\,\frac{1}{N}\Tr (W_{(n-2)}X_{n-1}X^\dagger_{n-1})^m 
 \eea
then using \eq{tCijW} and \eq{XWXcoeffs} we find the recursion relation
\bea
\frac{1}{N} \langle \Tr W_{(n-1)}^{m} \rangle_{W,n-1} &=&
\sum_{k=1}^m 
\sum_{\bm \in \cM_{m,k}}
\!\!\!\!\!\!  \tc_{1^m, 1^m} (m_1\,,\ldots ,m_k) 
\frac{1}{N} \langle \Tr W_{(n-2)}^{m_1}\rangle_{W,n-2}\ldots  \frac{1}{N}\langle\Tr W_{(n-2)}^{m_k} \rangle_{W,n-2}\,. \non \\
&& \label{MultiWishartRecursion}
\eea
Comparing \eq{MultiWishartRecursion} with \eq{ProdGinibreMoment} for $({\bf i,j})=(1^m,1^m)$ we see that
\bea
\frac{1}{N}\langle \Tr W_{(n)}^{m}\rangle_{W,n} &=& C^{(n)}_{1^m,1^m} \,. \label{WCequality}
\eea 
We note that \eq{WCequality} gives the equality of the following moments in the multi-Gaussian ensemble (see also the equality \eq{MomentEquality1})
\bea
\frac{1}{N}\langle \Tr (A_1A_1^\dagger \ldots A_n A_n^\dagger)^m \rangle &=&  \frac{1}{N} \langle \Tr ( A_1\ldots A_n A_n^\dagger \ldots A_1^\dagger) ^m \rangle\,.
\eea
Finally, from \eq{WCequality} and \eq{SingularMomentsFC} we have
\be
\frac{1}{N} \langle \Tr W_{(n)}^{m}\rangle_{W,n} = FC_{n}(m)\,. \label{WmnFC}
\ee
and in appendix \ref{App:MultiWishartMoments} we have checked \eq{WmnFC} explicitly for low values of $m$. Our final expression for the general mixed moment in the product Ginibre ensemble is the main result of this paper:
\be
\boxed{C^{(n)}_{\bf i,j} =\sum_{k=1}^m 
\sum_{\bm \in \cM_{m,k}}  
\tc_{{\bf i}, {\bf j}} (m_1\,,\ldots ,m_k)
FC_{n-1}(m_1) \ldots FC_{n-1}(m_k)} \label{ProdGinibreMomentFC}\,.
\ee

As a mathematical aside, we note that it follows from \eq{MultiWishartRecursion} and \eq{WmnFC} that the Fuss-Catalan numbers satisfy the following recursion relation\footnote{
In example 5. section 7.5 of \cite{ConcreteMathematics} we find that Fuss-Catalan numbers satisfy a similar but different recursion relation. Our recursion relation relates Fuss-Catalan numbers with different values of $n$ whereas the relation in \cite{ConcreteMathematics} closes for a given value of $n$.
}
\bea
 FC_{n-1}(m) &=&\sum_{k=1}^m 
\sum_{\bm \in \cM_{m,k}}
\tc_{1^m, 1^m} (m_1\,,\ldots ,m_k)
FC_{n-2}(m_1) \ldots  FC_{n-2}(m_k)  \label{FCRecursion}\,.
\eea 
with $\tc_{1^m, 1^m}$ given by \eq{XWXcoeffs}. 

\subsection{Examples}

The examples from section \ref{sec:Simplen2Examples} for general $n$ are verified in appendix \ref{App:MultiWishartMoments} when we check the recursion relation \eq{WmnFC}.  We can also compute the examples from sections \ref{sec:Example2} and \ref{sec:Example3} as follows (using the same diagrammatica as before):
\bea
\frac{1}{\sig^{2m}}C^{(n)}_{m,m}&=&\frac{1}{N} \langle \Tr X_{(n)}^m (X_{(n)}^\dagger )^m\rangle \\
&=& \Blp \frac{1}{N}\langle \Tr W_{n-1} \rangle_{W,n-1}\Brp ^m \\
&=& (FC_{n-1}(1))^m \\
&=& 1
\eea
and
\bea
\frac{1}{\sig^{2m}}C^{(n)}_{[m-1,1],[1,m-1]}&=& \frac{1}{N} \langle \Tr X_{(n)}^m X_{(n)}^\dagger X_{(n)} (X_{(n)}^\dagger )^m\rangle \\
&=&  \frac{1}{N^{m+1}}  \langle \Tr W_{n-1} \rangle_{W,n-1}^{m+1} + \frac{1}{N^{m}} \langle \Tr W_{n-1}^2 \rangle_{W,n-1}\langle \Tr W_{n-1} \rangle_{W,n-1}^{m-1}  \\
&=& FC_{n-1}(1)^{m+1} + FC_{n-1}(2) FC_{n-1}(1)^{m-1}  \\
&=& n+1 \label{SwithcMoments_n}
\eea
Our numerical experiments are summarized in tables \ref{table:NumericalExamplesSwitch_n} and give excellent agreement with \eq{SwithcMoments_n}.

\begin{table}[!htb!]
\begin{subtable}{.5\textwidth}
\centering
\begin{tabular}{c|ccccc}
 m & $\frac{1}{N}\langle \Tr X_{(3)}^m X_{(3)}^\dagger X_{(3)} (X_{(3)}^\dagger )^m\rangle$\\
\hline
2 &4.000 \\
3 &4.001 \\
4 &4.004 \\
\end{tabular}
\end{subtable}
\begin{subtable}{.5\textwidth}
\centering
\begin{tabular}{c|ccccc}
 m & $\frac{1}{N}\langle \Tr X_{(4)}^m X_{(4)}^\dagger X_{(4)} (X_{(4)}^\dagger )^m\rangle$ \\
\hline
2 &4.999 \\
3 &4.998 \\
4 &4.998 \\
\end{tabular}
\end{subtable}
   \captionsetup{justification=centering, margin=3cm}
\caption{Numerical values of $\frac{1}{N} \langle \Tr X_{(n)}^m X_{(n)}^\dagger X_{(n)} (X_{(n)}^\dagger )^m\rangle$
with $N=500$ and averaged from $1500$ samples.}
\label{table:NumericalExamplesSwitch_n}
\end{table}

\section{Discussion}

In this work we have studied the ensemble of products of Ginibre matrices. We have computed the probability distribution function and then the mixed moments, our main result being \eq{ProdGinibreMomentFC}. The slogan we have drawn from our investigation is that the product Ginibre ensemble remains Gaussian but not i.i.d. since the elements are no longer identically distributed. Another point of view is that the variance profile is randomly sampled from the Wishart ensemble; random couplings are certainly well known from models such a the Sherrington-Kirkpatrick model \cite{SherringtonKirkpatrick} and more recently the SYK model \cite{K2, Sachdev1993a, Maldacena:2016hyu} but the product Ginibre model is somewhat novel in that the random couplings are Wishart not Gaussian distributed.

There are several directions for future research, most pressing for the problems considered in this paper is the evaluation of mixed moments for complex Gaussian matrices with a deterministic variance profile. There are several closely related and slightly more general themes to explore, in particular the $\beta=1,4$ versions of the current work (which has $\beta=2$ in Dyson's classification). Certainly $\beta=1$ is necessary for applications to neural networks but our explorations so far indicate the results are essentially identical $\beta=2$. There is some rather more quantitative changes in the results for the mixed moments when considering rectangular matrices and it would be interesting to pursue this line of inquiry. 

Finally, one of our central goals was to understand the first subleading correction in $\frac{1}{N}$ to ensembles of products of matrices, we have only scratched the surface of this in section \ref{sec:FiniteN} and plan to look deeper into this. From our initial investigations in section \ref{sec:FiniteN}, it seems suggestive that while moments in the Ginibre ensemble are corrected only at order $\frac{1}{N^2}$, moments in the product Ginibre ensemble receives corrections at order $\frac{1}{N}$ coming from the expectation value of a multi-trace operator in the Wishart ensemble. This would be interesting to establish as such corrections are likely more straightforward than enumerating non-planar pairings in the Ginibre ensemble.

\vspace{1cm} \noindent {\bf Acknowledgements:} We would like to thank Jean-Bernard Zuber for collaboration at early stages of this project and numerous discussions throughout. NH would like also to thank Satya Majumdar for useful conversations,  Surya Ganguli for interesting discussions which stimulated his interest in products of random matrices and Ruth Corran for her patient explanations.
SL's work is supported by the Simons Foundation grant 488637 (Simons Collaboration
on the Non-perturbative bootstrap) and the project CERN/FIS-PAR/0019/2017. Centro
de Fisica do Porto is partially funded by the Foundation for Science and Technology of
Portugal (FCT) under the grant UID-04650-FCUP. 


\section*{Appendices}

\begin{appendix}

\section{Notation}

Our notation for various expectation values is as follows:
\bea
\langle \cO(X,X^\dagger)\rangle_G &=& \int d^2X \, \cO(X,X^\dagger) P(X,X^\dagger) \\
\langle \cO(X,X^\dagger)\rangle_{(n)} &=&  \int d^2X \, \cO(X,X^\dagger) P_{(n)}(X,X^\dagger) \\
\langle \cO(W)\rangle_{W} &=&  \int dW \, \cO(W) P_{W}(W) \\
\langle \cO(W_1,\ldots ,W_n)\rangle_{W,n} &=&  \int \prod_{i=1}^n dW_i \, \cO(W_1,\ldots ,W_n) P_{W,n}
\eea
where the probability distributions are defined in \eq{GinibreProbDefinition}, \eq{eq:ProbGinibre}, \eq{WLProbDefintiion}, \eq{eq:WishartnPDF}. The measure
$d^2X$ for a complex matrix $X$ is defined by
\be 
d^2X =dX\,dX^\dagger =\prod_{i,j=1}^N \Re\left(X_{ij}\right)\Im\left(X_{ij}\right)\,.
\ee
We also denote the integral over Hermitian positive-definite matrices $W$ by
\be
\int_{W>0}dW\,.
\ee
\section{Fuss-Catalan numbers}\label{app:FussCatalan}
There is a small amount of discrepancy with what are referred to as {\it Fuss-Catalan} numbers. There are references which refer to a three-parameter sequence as the Fuss-Catalan numbers, we will use the definition from section 7.5 of \cite{ConcreteMathematics} albeit with slightly different notation:
\bea
FC_{n}(m) = \frac{1}{nm+1} \binom{(n+1)m}{m}  \,.\label{FC_definition}
\eea
When $n=1$ these reduce to the Catalan numbers 
\be
FC_{1}(m)  = C_m\,.
\ee
The Fuss-Catalan numbers appear in numerous places in the text and we have derived a novel recursion relation which they satisfy \eq{FCRecursion}.

\section{Product of Gaussian scalars}

We include here an exposition of the ensemble obtained from the product of two complex Gaussian scalars $a$ and $b$ distributed in $\cN(0,1)$. This is a simple and well known computation but we include it here as it served to enlighten us on how to think about the probability distribution function of the product of Ginibre matrices.

Given the probability distribution function of a complex scalar in $\cN(0,1)$, \textit{viz.}
\be 
P(a,a^*)=\frac{1}{2\pi}\int d^2a\,e^{-\frac{1}{2}|a|^2}\,,
\ee
the probability distribution function for the product ensemble is 
\bea
P(z,z^*) &=& \frac{1}{4\pi^2} \int d^2a\, d^2b\, \delta^{(2)}(z-ab) e^{-|a|^2/2-|b|^2/2} \nonumber \\
&=&  \frac{1}{4\pi^4} \int d^2a \, d^2b\,d^2t \, e^{it(z-ab)+it^*(z-ab)^*} e^{-\frac{1}{2} |a|^2-\frac{1}{2} |b|^2} \nonumber \\
&=& \frac{1}{4\pi^4} \int d^2t \, e^{itz+it^*z^*}\int d^2a \, d^2b\,e^{-\frac{1}{2}(a - i 2t^* b^*)(a^*- i 2t b)  - 2 tt^* bb^*} e^{-\frac{1}{2}bb^*}\nonumber \\
&=& \frac{1}{\pi^2} \int d^2t \,\frac{e^{itz+it^*z^*}}{1+4\,tt^*}
\label{eq: scalar prod pdf}\,.
\eea
where we now see that the moment generating function $\phi_z(t,t^*)$ is given by
\bea
\phi_z(t,t^*) &=& \pi^2 \int d^2z \,
e^{-itz-it^*z^*}
P(z,z^*) =  \frac{1}{1+4tt^*}\,,
\eea
where we have used the normalization conventions of the main text.
Continuing on and using the Schwinger trick in \eqref{eq: scalar prod pdf}, we get
\bea
P(z,z^*)
&=& \frac{1}{\pi^2} \int d^2t \int_0^\infty du\, e^{itz+it^*z^*} e^{-u(1+4tt^*)}  \\
&=& \frac{1}{2\pi} \int du \frac{1}{u} e^{-u -\frac{zz^*}{4u}} 
\label{eq: scalar prod pdf 2}\\
&=& \frac{1}{2\pi} K_0(|z|^2)\,.
\eea
The reader may compare Equation \eqref{eq: scalar prod pdf 2} derived above with the
Equation \eqref{PDFX2} obtained in the main text, setting $N=1$ there.

\section{Multi-variable gamma functions}\label{app:MultiGamma}

One of the key inputs to our analysis is the expression in Equation
\eqref{eq: schwinger determinant}
for the determinant of
a Hermitian positive definite matrix in terms of an integral over Hermitian
positive definite matrices. This follows from the definition of the complex
multivariate Gamma function. In this appendix we will prove this relation
and provide some additional details about this function.

The complex multivariate Gamma function is given by \cite{james1964}
\bea\label{eq:cplx gamma}
\Gam_N(a) &=& \int_{W=W^\dagger >0} dW \det W^{a-N} e^{-\Tr W}\,,\\
&=& \label{eq: cplx gamma prod}
\pi^{N(N-1)/2}\prod_{k=1}^N\,\Gamma\left(a-k+1\right)\,,
\eea
where $\Gamma$ is the usual gamma function and $\Re(a)>N-1$.
The integration is over $N\times N$ Hermitian positive definite matrices $W$.

We will use Equation \eqref{eq:cplx gamma} 
to provide an integral representation for the 
determinant of a Hermitian positive definite matrix. This may usefully be thought
of as a multi-variable or matrix generalization of the Schwinger trick, \textit{viz.}
\bea\label{eq: std schwinger}
\Gamma(a) &=& \int_0^\infty du\,u^{a-1}\,e^{-u} =
\lambda^a\,\int_0^\infty dv\,v^{a-1}\,e^{-\lambda\,v}\nonumber \\
\Rightarrow
\frac{1}{\lambda^a} &=& \frac{1}{\Gamma(a)}\int_0^\infty dv\,
v^{a-1}\,e^{-\lambda\,v}\,,
\eea
where we defined $u=\lambda\,v$.

With Equation \eqref{eq: std schwinger} in mind, we now return to Equation
\eqref{eq:cplx gamma}. Consider a positive definite Hermitian matrix $\Sig$, 
it has a unique square root $\Sig_0$ which is also positive definite Hermitian:
\be
\Sig_0\Sig_0=\Sig\,.
\ee
As in the univariate case \eqref{eq: std schwinger}, we do the change of variables
\footnote{It is straightforward to check that $V$ is also hermitian positive
definite. It is quite straightforward to verify these relations when $\Sig$ is
proportional to the identity matrix, in which case
\bea
\Sig &=& b 1\!\!1\,,\quad \det \Sig = b^N\\
d(\Sig_0^{-1/2} V \Sig_0^{-1/2})  &=& b^{-N^2} dV = \det \Sig^{-N} dV\,.
\eea
}
\bea
W &=& \Sig_0\,V\Sig_0\,,
\qquad dW = \det W = \det \Sigma\,\det V\,,
\qquad \left(\det \Sig\right)^N\,dV\,,
\eea
in Equation \eqref{eq:cplx gamma} to obtain
\bea
\Gam_N(a) &=& \left(\det \Sig\right)^a
\int_{V=V^\dagger >0} dV \det V^{a-N} e^{-\Tr \Sig\,V}\,,
\eea
and hence
\be
\boxed{\frac{1}{\det \Sig^{a}} = \frac{1}{\Gam_N(a)} \int_{W >0} dW \det W^{a-N} e^{-\Tr \Sig\, W}}\,,
\ee
where we relabeled $V$ as $W$. The integral $ \int_{W >0} dW$ is the integral over the space of positive definite Hermitian $W$. This is the main identity we use in the text. It is
apparent that the $N=1$ case of this identity is Equation \eqref{eq: std schwinger}.
For this reason, we refer to this as the matrix generalization
of the Schwinger trick.

\section{Moment generating function for the product of multiple Gaussians}\label{App:MomGenMultiple}
In this Appendix we will prove the Equation \eqref{eq:ProbGinibre} 
in the main text for the probability distribution function  for the
product of $n$ Ginibre matrices by computing the moment generating function. 
We start with the probability distribution of the product
of $n$ Ginibre matrices in the form
\be
\begin{split}
P_{(n)}&(X,X^\dagger)=\left(\frac{N^{n}}{\pi^n\sig^2}\right)^{N^2} \int \cD^{(n)} A\, \delta(X -  \prod_{i=1}^n A_1) 
e^{-N \sum_{i=1}^n \frac{1}{\sig_i^2}\Tr A_iA_i^\dagger} \\
&= 
 \frac{1}{\pi^{2N^2}}\left(\frac{N^{n}}{\pi^n\sig^2}\right)^{N^2} \int d^2T e^{i\Tr TX + i \Tr T^\dagger X^\dagger} \int \cD^{(n)} A\,  e^{-i \Tr  TA -i \Tr (TA)^\dagger } 
e^{-N \sum_{i=1}^n \frac{1}{\sig_i^2}\Tr A_iA_i^\dagger}\,.
\end{split}
\ee
Therefore the moment generating function is
\bea
\vphi_{(n)}(T,T^\dagger)&=&\left(\frac{N^{n}}{\pi^n\sig^2}\right)^{N^2} \int \cD^{(n)} A\,  e^{-i \Tr  TA -i \Tr (TA)^\dagger } 
e^{-N \sum_{i=1}^n \frac{1}{\sig_i^2}\Tr A_iA_i^\dagger}\,.
\eea
We will first integrate out $A_n$, then $A_{n-1}$, and so on, finally integrating
out $A_1$. Doing the $A_n$ and $A_{n-1}$ integrals is straightforward, and follows
the $n=2$ computations. We eventually find
\footnote{We have defined the quantity $|M|^2=M^\dagger M=M\,M^\dagger$ in the
determinant for a complex matrix $M$. This is
\textit{a priori} not well-defined as $M$ and $M^\dagger$ don't commute. However,
under the determinant
\be 
\det\left(1+M^\dagger M\right)=e^{\tr\ln\left(1+M^\dagger M\right)}
=e^{\tr\ln\left(1+M\,M^\dagger\right)}=\det\left(1+M\,M^\dagger\right)\,,
\ee
as we may readily verify from the series expansion of $\ln$. 
Hence $\det\left(1+|M|^2\right)$, which is what we work with, is indeed well defined.
}
\bea
\vphi_{(n)}(T,T^\dagger)&=& 
\left(\frac{N^{n}}{\pi^n\sig^2}\right)^{N^2}  \left(\frac{\pi^2(\sig_n\sig_{n-1})^2}{N^2}\right)^{N^2}  \int \cD^{(n-2)} A\,  
e^{-N \sum_{i=1}^{n-2} \frac{1}{\sig_i^2}\Tr A_iA_i^\dagger } \non\\&&\qquad\qquad
\frac{1}{\det\Blp1\!\!1+ \frac{(\sig_{n-1}\sig_n)^2}{N^2} |TA_1...A_{n-2}|^2\Brp^{N}} \non\\
&=&
\left(\frac{N^{n}}{\pi^n\sig^2}\right)^{N^2} \left(\frac{\pi^2(\sig_n\sig_{n-1})^2}{N^2}\right)^{N^2}  
\int \cD^{(n-2 )} A\,  e^{-N \sum_{i=1}^{n-2} \frac{1}{\sig_i^2}\Tr A_iA_i^\dagger } 
\non\\&&\qquad\qquad
\int_{W_1>0} dW_1 e^{-\Tr W_1} e^{- \frac{(\sig_{n-1}\sig_n)^2}{N^2} \Tr |TA_1...A_{n-2}|^2 W_1}\,.
\eea
where $|TA_1...A_{n-2}|^2=TA_1\ldots A_{n-2}A_{n-2}^\dagger\ldots 
A_1^\dagger T^\dagger$. We can now integrate out $A_{n-2}$ and beyond. It is
useful to write
\bea
\Tr |TA_1...A_{n-2}|^2 W_1
&=& \Tr A_{n-2}A_{n-2}^\dagger | W_1^{1/2}TA_1... A_{n-3}|^2 \,.
\eea
We may now integrate out $A_{n-2}$ to find
\bea
\vphi_{(n)}(T,T^\dagger)&=&
\frac{1}{\pi^{2N^2}}  \frac{N^{nN^2}}{\sig^{2N^2}}  \left(\frac{(\sig_n\sig_{n-1}\sig_{n-2})^2}{N^3}\right)^{N^2}  
\int \cD^{(n-3)} A\,  e^{-N \sum_{i=1}^{n-3} \frac{1}{\sig_i^2}\Tr A_iA_i^\dagger }  
\\
&& \int_{W_1>0} dW_1 e^{-\Tr W_1}  
\frac{1}{\det\Blp1\!\!1 + \frac{(\sig_1\sig_2\sig_3)^2}{N^3} |W_1^{1/2} TA_1...A_{n-3}|^2  \Brp^N}\\
&=&
\frac{1}{\pi^{2N^2}}  \frac{N^{nN^2}}{\sig^{2N^2}}  \left(\frac{(\sig_n\sig_{n-1}\sig_{n-2})^2}{N^2}\right)^{N^2}  \int \cD^{(n)} A\,  
e^{-N \sum_{i=1}^{n-3} \frac{1}{\sig_i^2}\Tr A_iA_i^\dagger }  \\
&& \int_{W_i>0} dW_1 dW_{2} e^{-\sum_{i=1}^{2}W_i} e^{- \frac{(\sig_n\sig_{n-1}\sig_{n-2})^2}{N^3}\Tr  |W_2^{1/2}W_1^{1/2} TA_1...A_{n-3}|^2 }\,.
\eea
We may similarly integrate out all $A_i$ to get
\bea
\vphi_{(n)}(T,T^\dagger)&=& \frac{1}{\Gam_N(N)^{n-1}} \int_{W_i>0} \prod_{i=1}^{n-1}dW_i e^{-\sum_{i=1}^{n-1}W_i} 
e^{- \frac{\sig^2}{N^{n} } \Tr TT^\dagger W_1W_2\ldots W_{n-1}}\,.
\eea
Finally we should rescale the $W_i$ by $N$ such that the action is $NW_i$:
\bea
\vphi_{(n)}(T,T^\dagger)&=& \frac{N^{(n-1)N^2}}{\Gam_N(N)^{n-1}} \int_{W_i>0} \prod_{i=1}^{n-1}dW_i e^{-N\sum_{i=1}^{n-1} \Tr W_i} 
e^{- \frac{\sig^2}{N } \Tr \, TT^\dagger W_1W_2\ldots W_{n-1}}\,.
\eea
It is straightforward to Fourier transform this expression and obtain $P_{(n)}$ in
eq \eqref{eq:ProbGinibre}.

\section{Exact computation of $ \Tr W_{(n)}^{m} $}\label{App:MultiWishartMoments}
In this section we will explicitly verify the expression \eq{WmnFC} for the
moments $\Tr W_{(n)}^{m}$ for some low values of $m$. For the reader's convenience,
we reproduce \eq{WmnFC} here
\be
\frac{1}{N} \langle \Tr W_{(n)}^{m}\rangle = FC_{n}(m)\,.
\ee
We will use the recursion relation \eq{MultiWishartRecursion}
\be
\frac{1}{N} \langle \Tr W_{(n-1)}^{m} \rangle_{W,n-1} =
\sum_{k=1}^m 
\sum_{\bm \in \cM_{m,k}}
\!\!\!\!\!\!  \tc_{1^m, 1^m} (m_1\,,\ldots ,m_k) 
\frac{1}{N} \langle \Tr W_{(n-2)}^{m_1}\rangle_{W,n-2}\ldots  \frac{1}{N}\langle\Tr W_{(n-2)}^{m_k} \rangle_{W,n-2}\,,
\ee
alongwith the coefficients \eq{XWXcoeffs}. 
We will also need the boundary condition
\be
\langle W_{(1)}^m\rangle = C_m\,,
\ee
where $C_m$ is the Catalan number.
\vskip 5mm
\noindent $ \mathbf{\Tr W_{(n-1)}}$ \\
The first moment is simple, the only contribution to the recursion relation is
\be 
\tc_{1,1}(1)=1\,.
\ee
Hence
\be 
\frac{1}{N} \langle \Tr W_{(n-1)}^{m} \rangle_{W,n-1} =
\frac{1}{N} \langle \Tr W_{(n-2)}^{m} \rangle_{W,n-2} =
\frac{1}{N} \langle \Tr W_{(1)}^{m} \rangle_{W,1} = 1\,,
\ee
where we have repeatedly used the first equality to arrive at the second, and
subsequently used $C_1=1$. Therefore
\bea
\frac{1}{N} \langle \Tr W_{(n-1)}^{m} \rangle_{W,n-1}
&=& 1 \\
&=& FC_{n-1}(1)\,.
\eea

\vskip 5mm
\noindent $ \mathbf{\Tr  W_{(n-1)}^2}$ \\
We first recall that
\be
\tc_{1^2,1^2}(2)= \tc_{1^2,1^2}(1,1) = 1
\ee
and 
\be
\langle \Tr W_{(1)}^2 \rangle = 2
\ee
then 
\bea
\langle \Tr W_{(n-1)}^2 \rangle &=& \langle \Tr W_{(n-2)}^2 \rangle + 1 \\
&=& n \\
&=& FC_{n-1}(2)\,.
\eea

\vskip 5mm
\noindent $ \mathbf{\Tr W_{(n-1)}^3}$ \\
We have
\bea
\tc_{1^3,1^3}(1,1,1) &=& 1\,,\quad \tc_{1^3,1^3}(1,2)=3\,,\quad \tc_{1^3,1^3}(3)=1
\eea
then 
\bea
\langle \Tr  W_{(n-1)}^3 \rangle &=&\langle \Tr  W_{(n-2)}^3 \rangle + 3  FC_{n-2}(2) +1
\eea
which is solved by
\bea
\langle \Tr  W_{(n-1)}^3 \rangle &=& FC_{n-1}(3)\,.
\eea

\vskip 5mm
\noindent $ \mathbf{\Tr W_{(n-1)}^4}$ \\
We have
\bea
\tc_{1^4,1^4}(1,1,1,1) = &1&\,,\quad \tc_{1^4,1^4}(1,1,2)=5\,,\quad 
\tc_{1^4,1^4}(1,3)=4\,,\quad \non\\
\tc_{1^4,1^4}(2,2)&=&2\,,\quad 
\tc_{1^4,1^4}(4)=1\,.
\eea
As a result,
\be
\langle \Tr  W_{(n-1)}^4 \rangle =\langle \Tr  W_{(n-2)}^4 \rangle +2FC_{n-2}(2)^2
+   FC_{n-2}(4)+4FC_{n-2}(3)+6FC_{n-2}(2) +1\,,
\ee
which is solved by
\bea
\langle \Tr  W_{(n-1)}^3 \rangle &=& FC_{n-1}(4)\,.
\eea

\vskip 5mm
\noindent $ \mathbf{\Tr W_{(n-1)}^5}$ \\
Starting with
\bea
&& \tc_{1^5,1^5}(1,1,1,1,1) = 1\,,\quad \tc_{1^5,1^5}(1,1,1,2)=10\,,\quad \tc_{1^5,1^5}(1,1,3)=10\,, \\
&&  \tc_{1^5,1^5}(1,4)=5\,,\quad \tc_{1^5,1^5}(1,2,2)=10\,,\quad \tc_{1^5,1^5}(2,3)=5\,,\quad \tc_{1^5,1^5}(5)=1\,. 
\eea
we have
\bea
\langle \Tr (W_{(n-1)})^5 \rangle 
&=&\langle  \Tr W_{(n-2)}^5 \rangle + 10  FC_{n-2}(2) + 10 FC_{n-2}(3) +  5 FC_{n-2}(4) \\
&& + 10 FC_{n-2}(2)^2 + 5 FC_{n-2}(2)FC_{n-2}(3) +1 
\eea
which is solved by
\bea
\langle  \Tr W_{(n-1)}^5 \rangle &=& FC_{n-1}(5)\,.
\eea

\end{appendix}

\providecommand{\href}[2]{#2}\begingroup\raggedright\endgroup

\end{document}